\documentclass[a4paper,11pt]{article}
\pdfoutput=1 
\usepackage[symbol]{footmisc} 
\usepackage{jheppub} 

\usepackage[T1]{fontenc} 
\usepackage[utf8]{inputenc}
\usepackage[T1]{fontenc}
\usepackage{epsfig,latexsym}
\usepackage{amsmath}
\usepackage[dvipsnames]{xcolor}
\usepackage{adjustbox}
\usepackage{graphicx}
\usepackage{verbatim}
\usepackage{mathrsfs}
\usepackage{amssymb}
\usepackage{multirow}
\usepackage{epsfig}
\usepackage{color,colordvi}
\usepackage{appendix}
\usepackage{slashed}
\usepackage{cancel}
\usepackage{float}
\usepackage{caption}
\usepackage{epsf}
\usepackage{amsmath}
\usepackage{physics}
\usepackage{MnSymbol}
\usepackage[normalem]{ulem}
\DeclareMathAlphabet{\mathpzc}{OT1}{pzc}{m}{it}

 \csname
@addtoreset\endcsname{equation}{section}

\newcommand{\beq}{\begin{equation}}
\newcommand{\eeq}{\end{equation}}
\renewcommand{\[}{\left[}
\renewcommand{\]}{\right]}
\renewcommand{\(}{\left(}
\renewcommand{\)}{\right)}
\def\eq#1{{Eq.~(\ref{#1})}}
\def\eqs#1#2{{Eqs.~(\ref{#1})--(\ref{#2})}}
\def\fig#1{{Fig.~\ref{#1}}}

\def\sect#1{{Sect.~\ref{#1}}}

\def\be{\begin{equation}}
\def\ee{\end{equation}}
\def\bes{\begin{equation*}}
\def\ees{\end{equation*}}
\def\bead{\begin{aligned}}
\def\eead{\end{aligned}}
\def\bmat{\left(\begin{matrix}}
\def\emat{\end{matrix}\right)}
\def\cA{{\cal A}}
\def\cL{{\cal L}}
\def\cC{{\cal C}}

\def\cO{{\cal O}}

\title{The Anomalous Case of Axion EFTs and Massive Chiral Gauge Fields}

\author[a]{Quentin~Bonnefoy,}
\author[a]{Luca~Di~Luzio,}
\author[a,b]{Christophe~Grojean,}
\author[a,b]{Ayan~Paul}
\author[a,b]{and Alejo~N.~Rossia}

\affiliation[a]{DESY, Notkestra{\ss}e 85, D-22607 Hamburg, Germany}
\affiliation[b]{Institut f\"ur Physik, Humboldt-Universit\"at zu Berlin, D-12489 Berlin, Germany}

\emailAdd{quentin.bonnefoy@desy.de}
\emailAdd{luca.diluzio@desy.de}
\emailAdd{christophe.grojean@desy.de}
\emailAdd{ayan.paul@desy.de}
\emailAdd{alejo.rossia@desy.de}

\abstract{We study axion effective field theories (EFTs), with a focus on axion couplings to massive chiral gauge fields. We investigate the EFT interactions that participate in processes with an axion and two gauge bosons, and we show that, when massive chiral gauge fields are present, such interactions do not entirely originate from the usual anomalous EFT terms. We illustrate this both at the EFT level and by matching to UV-complete theories. In order to assess the consistency of the Peccei--Quinn (PQ) anomaly matching, it is useful to introduce an auxiliary, non-dynamical gauge field associated to the PQ symmetry. When applied to the case of the Standard Model (SM) electroweak sector, our results imply that anomaly-based sum rules between EFT interactions are violated when chiral matter is integrated out, which constitutes a smoking gun of the latter.  As an illustration, we study a UV-complete chiral extension of the SM, containing an axion arising from an extended Higgs sector and heavy fermionic matter that obtains most of its mass by coupling to the Higgs doublets. We assess the viability of such a SM extension through electroweak precision tests, bounds on Higgs rates and direct searches for heavy charged matter. At energies below the mass of the new chiral fermions, the model matches onto an EFT where the electroweak gauge symmetry is non-linearly realised.}

\begin{document} 
\begin{flushright}
DESY 20-194\\
HU-EP-20/32
\end{flushright}
\maketitle
\flushbottom

\section{Introduction}

Axions are prime and ubiquitous candidates for physics beyond the Standard Model (BSM), and as such have motivated great theoretical and experimental efforts. Starting with the Peccei--Quinn (PQ) solution to the strong CP problem \cite{Peccei:1977hh,Peccei:1977ur} and the prediction of the associated QCD axion \cite{Weinberg:1977ma,Wilczek:1977pj}, numerous axion\footnote{Henceforth, we use the word axion to refer to any axion-like particle (ALP), and call ``QCD axion'' an axion which is designed to solve the strong CP problem. We also refer to the axion shift symmetry as a PQ symmetry, irrespective of the presence of a coupling to gluons.} models have been built to address various formal or phenomenological questions (see \cite{DiLuzio:2020wdo} for a recent review). Moreover, based on astrophysical, cosmological and collider data, fair portions of their parameter spaces have been probed already. 

To extract the full information contained in the current experimental bounds, or to prepare the ground for a hypothetical observation of an axion and its couplings, we must identify the relevant set of parameters that best describe the axion phenomenology. Due to the mass gap between the axion and the rest of the UV sector in common axion models, this is usually done in terms of an axion effective field theory (EFT), where only the axion is present and interacts via non-renormalizable operators with other light particles (usually, SM fields). An axion EFT then contains all the necessary and sufficient operators to capture the axion phenomenology, and it allows one to obtain robust UV-independent bounds. In this context, two very relevant questions are: {\it (i)} what is the most general shape of the axion EFT? {\it (ii)} given an observation of the axion couplings that would fix some parameters in the axion EFT, what can be inferred about the UV?

The first question can be answered by systematically building all EFT operators allowed by the IR symmetries and the set of IR degrees of freedom (d.o.f.s) \cite{Georgi:1986df,Brivio:2017ije}. Yet, as suggested by question {\it (ii)}, we would also like to get a clear picture of the match between UV models and their IR EFTs, in order to, for instance, map specific EFTs structures to UV properties. This can be addressed by studying the explicit matching between UV theories and the relevant low-energy axion EFTs, either in full generality as in \cite{Srednicki:1985xd}, or by studying precise and well-motivated models \cite{Kim:1979if,Shifman:1979if,Dine:1981rt,Zhitnitsky:1980tq}. 

Within the EFT, the set of couplings involving the axion and two gauge fields is of particular interest. Among those, the couplings to massless SM vector bosons, namely photons and gluons, have been extensively studied. This is justified by the theoretical and phenomenological interest for the QCD axion, whose coupling to gluons solves the strong-CP problem and whose decay to photons is generically the only kinematically allowed one. The photon coupling is also the most relevant EFT coupling in several astrophysical media, as well as in most current experimental searches.

When studying how the coupling to photons or gluons arises in specific UV-complete models, one obtains the striking result that they are expressed in terms of the mixed anomalies between the PQ symmetry and the relevant gauge symmetry in the UV \cite{Steinberger:1949wx}. Those anomalies do not run under the RG flow and yield, in the IR, transparent information about the UV PQ breaking sector, providing a first answer to question {\it (ii)}. It is also quite fortunate that such couplings can be inferred from anomalies only, since they can therefore be transported from a weakly-coupled UV theory down to the IR, even in the presence of non-perturbative physics \cite{Wess:1971yu,Witten:1983tw}.

On the other hand, the couplings to the $Z$ or $W$ bosons are quite irrelevant for QCD axion physics; they do not affect astrophysical or cosmological studies while their impact at colliders is negligible. Consequently, less attention has been devoted to axion couplings with those massive gauge bosons\footnote{See \cite{Alonso-Alvarez:2018irt} for the model-independent contributions to the QCD axion-electroweak bosons couplings, as well as \cite{Mimasu:2014nea,Jaeckel:2015jla,Brivio:2017ije,Bauer:2017nlg,Bauer:2017ris,Mariotti:2017vtv,Bauer:2018uxu,CidVidal:2018blh,Gavela:2019cmq,Coelho:2020saz} for collider searches and \cite{Batell:2009jf,Izaguirre:2016dfi,Gavela:2019wzg,Bauer:2019gfk,Cornella:2019uxs,Calibbi:2020jvd} for flavour physics studies.}. Nevertheless, one may expect that their study would also yield interesting IR and UV information: $Z$ and $W$ bosons are associated to a constraining $SU(2)_L\times U(1)_Y$ gauge symmetry, where we denote the gauge group of the SM as $SU(3)_C\times SU(2)_L\times U(1)_Y$, which for instance imposes that mixed PQ anomalies have a particular structure in UV models. If the axion couplings to the $Z$ or $W$ bosons were again given by mixed PQ anomalies, the IR phenomenology of the axion would be constrained, answering question {\it (i)}. Taking into account the aforementioned direct UV origin of anomalies, that would give IR constraints before an observation is made and UV information after. 

However, the fact that the $Z$ and $W$ bosons are massive invalidates the conclusions drawn from the case of photons or gluons. In particular, there exist more EFT operators involving massive gauge bosons than massless ones, due to the possibility that some symmetries are realized non-linearly in the EFT \cite{Brivio:2017ije}. This resonates with recent computations~\cite{Quevillon:2019zrd,Quevillon:2020hmx} which showed that, in a given UV model, mixed anomalies between the PQ and the gauge symmetries do not control uniquely the couplings between the axion and massive chiral gauge fields, even at leading order when all the heavy degrees of freedom lie at arbitrarily high energy.

Therefore, it seems that questions {\it (i)} and {\it (ii)} remain open. We reformulate them into four more specific ones: 
\begin{enumerate}
\vspace{-5pt}
\item\label{anomMatch} Are the coefficients that couple an axion to two gauge fields always captured by a suitable UV PQ anomaly coefficient? If not, how are the UV anomalies represented in the axion EFT? 
\vspace{-7pt}
\item\label{generalEFT}  What is the most general EFT of an axion and massive gauge fields, and how is it UV completed? 
\vspace{-7pt}
\item\label{sumRules} Can we tell different UV theories apart when probing axion couplings to SM massive gauge bosons at low-energies?
\vspace{-7pt}
\item\label{SMcase} Are there phenomenologically relevant and viable models that realize the most general EFT of an axion and electroweak gauge bosons?
\vspace{-5pt}
\end{enumerate}
Those are the questions we answer in this paper, which is organized as follows.

In section \ref{chiralMatterSection}, we start by introducing the notion of chirality for massive fermions that we repeatedly use in this paper. In short, we henceforth call a massive fermion chiral when its left- and right-handed components do not couple identically to some of the gauge fields. 

Then, in section \ref{physicalEFTsection}, we address questions \ref{anomMatch} and \ref{generalEFT}, in the simple case of an abelian gauge theory. We show that the axion coupling to the gauge field must correspond to a UV PQ anomaly coefficient when the gauge field is massless, whereas there exist additional EFT coefficients when the gauge field is massive, that are not constrained by anomaly matching but contribute to the amplitude connecting an axion and two vector bosons. Although these statements can be made at the EFT level already, we confirm them by an explicit matching to simple UV theories. It allows us to show that the additional, non-anomalous coefficients are only generated when there exist heavy fermions with chiral charges under the gauge symmetry. We also discuss the conditions under which the leading order axion-vector bosons amplitude is captured by the UV PQ anomaly coefficient{: for a coupling to two given (massless or massive) gauge fields, there always exists a choice of PQ charges in the UV which identifies the amplitude with the anomaly coefficient. However, this identification is not always possible for the axion couplings to all pairs of (massive) gauge fields simultaneously, since the different choices of PQ charges can be incompatible}. Finally, we find insightful to introduce an auxiliary PQ gauge field. It allows to identify anomalous terms made out of gauge fields only, called Generalized Chern-Simons (GCS) terms, which bridge the gap between UV PQ anomalies and EFT axion couplings, shedding a new light on our results.

We extend our analysis in section \ref{section:axionSM} to the case of a non-abelian gauge theory. We point out that non-anomalous operators also exist there, and we discuss them in detail for the specific case of the SM electroweak sector. We find that they allow to evade phenomenological correlations, which we express as sum-rules among Wilson coefficients of axion couplings to vector bosons, and which are linked to the structure of the mixed PQ anomalies with the $SU(2)_L\times U(1)_Y$ gauge symmetry. A violation of those sum-rules clearly indicates that the UV-completion of the axion EFT contains heavy fermions chiral with respect to the SM gauge group, answering question \ref{sumRules}.

Finally, in section \ref{eq:chiralextSM}, we address question \ref{SMcase}, namely we study a phenomenologically viable extension of the SM with heavy chiral fermion fields as well as an axion emerging from an extended Higgs sector (see Ref.~\cite{Bizot:2015zaa} for an attempt to classify the chiral extensions of the SM compatible with the measured Higgs rates). We present the low-energy EFT below the mass of the heavy fermions, and verify that the axion couplings to the photon, $Z$ and $W$ bosons violate the sum rules that would hold if those couplings were given by the UV PQ anomaly coefficients. Then, we show the compatibility of the model with current experimental data, in particular electroweak precision tests, bounds on Higgs rates and direct searches for heavy charged matter. Remarkably, such chiral extensions are still viable, although direct searches push them at the boundary of perturbativity.

Section \ref{conclusions} presents our conclusions. Two appendices complete this paper. First, appendix \ref{appendixMoreGaugeFields} completes and extends section \ref{physicalEFTsection}: there, we discuss the couplings of an axion to several abelian gauge fields, and we present explicitly the one-loop matching between simple UV theories with chiral fermions and the low-energy axion EFT. Second, appendix~\ref{appendixPheno} further discusses the axion couplings, the UV PQ anomalies, as well as their relation, in the model of section \ref{eq:chiralextSM}.
\section{Chirality and chiral extensions of the SM}\label{chiralMatterSection}

Chirality and anomalies are often intimately linked. As announced in the introduction, the axion EFT couplings to massive gauge fields will cease to be fixed by the mixed PQ gauge anomalies when \textit{chiral} matter is integrating out. It is notoriously difficult to build a SM extension involving new fermions forming a chiral and anomaly-free representation of electroweak symmetry, especially after the measurement of the Higgs boson production rate that forbids the simplest option of a fourth generation of quarks and leptons (see Ref.~\cite{Bizot:2015zaa} for an attempt to systematically identify such chiral models). However, the anomaly-free condition only imposes restriction on the charges of the particles in the UV but leaves open the possibility that particles belonging to a vector-like representation of $SU(2)_L\times U(1)_Y$ actually acquire a \textit{chiral} mass spectrum in the IR. In particular, it is perfectly possible
to get a collections of heavy fermions  whose left- and right-handed components do not couple symmetrically to the $W$ and $Z$, even if, globally, for each right-handed (RH) fermion, there exists a left-handed (LH) fermion with the same coupling to the $W$ and $Z$. Indeed, these paired fermions do not have to be linked by a mass term.  
In other words, a set of Weyl fermions whose gauge charges would be called vector-like but which are paired in a chiral fashion to form massive charged Dirac fermions would automatically be gauge-anomaly free. 

A simple example in the case of an abelian $U(1)$ symmetry is the following: consider four Weyl fermions with charges
\be
\psi_L\ : +1\ , \quad \psi'_L\ : 0\ , \quad \psi_R\ : 0\ , \quad \psi'_R\ : +1\ ,
\ee
where the subscripts indicate their chirality. Since there exists, for each left-handed fermion, a right-handed one with the same charge, the above spectrum would be called vector-like, but let us assume that the mass terms in the Lagrangian are as follows,
\be
\cL \supset -y\overline{\psi_L}\psi_R\phi-y'\overline{\psi'_L}\psi'_R\phi^\dagger+h.c. \supset -m\overline{\psi}\psi-m'\overline{\psi'}\psi' \ ,
\label{chiralExample}
\ee
where $\phi$ is a scalar field of charge $+1$ and vev $\frac{v}{\sqrt 2}$, and we defined the Dirac fermions $\psi={\tiny\left(\begin{matrix}\psi_L\\\psi_R\end{matrix}\right)},\psi'={\tiny\left(\begin{matrix}\psi'_L\\\psi'_R\end{matrix}\right)}$ with masses $m=y\frac{v}{\sqrt 2},m'=y'\frac{v}{\sqrt 2}$. The gauge invariant bare mass terms $\overline{\psi_L}\psi'_R+h.c.$, $\overline{\psi'_L}\psi_R+h.c.$ that are not included in \eqref{chiralExample} may be forbidden by global or discrete symmetries, as we discuss in Section~\ref{section:minimalChiralBSM} (vector-like masses will not change the low energy physics as long as they remain much smaller than the \textit{chiral} masses $m,m'$). As a result, the massive fermions $\psi,\psi'$ have different gauge charges when projected onto their left- or right-handed components, and they yield an example of what we refer to as chiral matter. Note that, due to the presence of the scalar $\phi$, the gauge theory is broken and the abelian field is massive.

Let us anticipate what follows and make the following remarks. As we will extensively discuss in section \ref{physicalEFTsection}, when integrated out, such fermions generate GCS interactions for the massive $U(1)$ gauge field and the would-be gauge field associated to the PQ symmetry, would it be gauged. These interactions cope for the PQ anomaly mismatch borne by the axion-massive photon interactions. Such a construction can be easily generalised to the full non-abelian $SU(2)_L\times U(1)_Y$ symmetry of the SM. In that case, such \textit{chiral} fermions give rise to an axion EFT Lagrangian where the $SU(2)_L\times U(1)_Y$ symmetry is non-linearly realised, as we discuss in section \ref{section:axionSM}.
In particular, if the Yukawa couplings of the new fermions are large enough compared to the SM gauge couplings, there is an interesting range of energy where the physics will be described by an EFT including all the SM degrees of freedom and the axion.
The chiral nature of the heavy fermions will then favour a HEFT-like Lagrangian rather a SMEFT-like one~\cite{Cohen:2020xca}.
\section{Axion couplings to gauge fields: the abelian case}\label{physicalEFTsection}

In this section, we discuss the (non-)anomalous properties of EFT operators that couple an axion to two gauge fields\footnote{We should emphasize that the role of axion terms in anomaly matching and their interplay with gauge field couplings are known facts in the literature, see for instance Ref.~\cite{Anastasopoulos:2006cz} and references therein. Nevertheless, such discussions usually concern models of additional broken gauge symmetries, where axions play the role of longitudinal components of massive gauge fields, while they have not been systematically applied to models of axions. Therefore, we detail here the important aspects and insist on the treatment and consequences specific to axion models.}, insisting on the case of massive chiral gauge fields and restricting ourselves to weakly coupled UV completions. We focus, to begin with, on a single abelian gauge field $A$, both for simplicity and because it already possesses most of the features that we wish to comment on. For instance, we restrict henceforth our discussion to processes involving one axion and two gauge fields, which are fully captured by the abelian formulae. We generalize our discussion to several abelian gauge fields in appendix \ref{appendixMoreGaugeFields} and, in the next section, we turn to phenomenologically relevant non-abelian theories by studying axion couplings to SM electroweak gauge fields.

\subsection{Non-anomalous EFT terms}\label{physicalEFTsection:abelianIntro}

Let us first consider axion couplings in the IR, namely using an EFT where the only remaining part of the PQ breaking sector is the pseudoscalar axion $a$. It couples to a light sector, which contains a gauge field in particular. We want to write down the CP-conserving theory of our single abelian gauge field and $a$ (henceforth, the CP-oddness of the axion and the CP-invariance of the action are always assumed). What are the lowest-dimensional EFT operators that mediate a coupling between the axion and the gauge field? At dimension 5, the answer is unique up to integration by parts, and well known: it is the ``$aF\tilde F$'' term,
\be
\cL_\text{EFT} \supset -g^2\frac{\cC}{16\pi^2f}a\, F_{\mu\nu}\tilde F^{\mu\nu} \ ,
\label{operator1}
\ee
where
$F_{\mu\nu}\equiv 2\partial_{[\mu}A_{\nu]}$  is the gauge field strength, $g$ is the gauge coupling, $\tilde F^{\mu\nu}\equiv\frac{\epsilon^{\mu\nu\rho\sigma}}{2}F_{\rho\sigma}$, $f$ is a dimensionful scale and $\cC$ an order one number. The axion is understood as the Nambu-Goldstone boson (GB) of a spontaneously broken PQ symmetry $U(1)_\text{PQ}$, whose action can be normalized such that $\delta_\text{PQ}a=\epsilon_\text{PQ}f$, the gauge fields being uncharged. The axion is not charged under the gauge symmetry $U(1)_A$ under which the gauge field shifts as $\delta_A A_\mu=\frac{1}{g}\partial_\mu\epsilon_A$.

Assuming an observation of this leading axion-gauge field effective interaction, one may like to deduce from it something about the UV theory that takes over \eqref{operator1} at high energies. There, there are particles charged under $U(1)_\text{PQ}$, and the usual statement is that the coefficient $\cC$ is the PQ anomaly coefficient $D^{\text{PQ}AA}$ of the UV theory\footnote{Namely,
\bes
D^{\text{PQ}AA}\equiv\sum_{\text{LH fermions }\psi}q^\text{PQ}_\psi (q^A_\psi){}^2-\sum_{\text{RH fermions }\psi}q^\text{PQ}_\psi (q^A_\psi){}^2 \ ,
\ees
with $q^\text{PQ}$ and $q^A$ the PQ and gauge charges respectively.}. Indeed, \eqref{operator1} shifts under the PQ symmetry,
\be
\delta_\text{PQ}\left(-g^2\frac{\cC}{16\pi^2f}a\, F_{\mu\nu}\tilde F^{\mu\nu}\right)=-g^2\epsilon_\text{PQ}\frac{\cC}{16\pi^2}\, F_{\mu\nu}\tilde F^{\mu\nu} \ ,
\label{anomalousShiftAxion}
\ee
and this shift has the correct form to be the anomalous variation of an effective action. In addition, we know that the EFT keeps track of the anomalies of any heavy fermionic UV sector that one would have integrated out. This is easily understood in the case of gauge theories with a spectrum split between heavy and light fermions; if the full spectrum was anomaly-free, yielding a consistent gauge theory, any low-energy EFT derived from it should also be consistent. Earlier studies \cite{DHoker:1984izu,DHoker:1984mif,Callan:1984sa,Paranjape:1985sk,Feruglio:1992fp,Masiero:1992wd,Anastasopoulos:2006cz} indeed showed that the consistency of the EFT is ensured by non-decoupling Wess-Zumino terms. For our axion EFT, the only remaining part of the heavy PQ-charged sector at low energies is the axion, whose coupling to gauge fields is fully captured by \eqref{operator1}.

Quevillon and Smith \cite{Quevillon:2019zrd,Quevillon:2020hmx} recently investigated the one-loop matrix elements between an axion and chiral gauge fields in simple UV models, and showed that they are not simply captured by the PQ anomaly coefficient of the UV theory. In our EFT language, they found that generically $\cC\neq D^{\text{PQ}AA}$, yielding an apparent contradiction with anomaly matching. Their explanations are phrased in terms of UV models 
{, which we discuss in the next section. Here, let us simply anticipate that the $U(1)_\text{PQ}$ symmetry to be considered in the UV is not always uniquely defined: there could exist a freedom to mix the would-be $U(1)_\text{PQ}$ with other global (vector-like) symmetries while keeping $\delta_\text{PQ} a=\epsilon_\text{PQ} f$, leaving an ambiguity in properly defining the PQ charges of the fermions contributing to the mixed $D^{\text{PQ}AA}$ anomaly. So a more precise question is rather whether it is possible to assign PQ charges to the UV degrees of freedom such that the coefficient $\mathcal{C}$ in \eqref{operator1} is given by the corresponding anomaly. In any case, let us simply notice that}
we can rewrite \eqref{operator1} so that it does not contribute to the PQ shift of the action anymore, while equally contributing to the physical amplitudes. For that, we can integrate it by parts to obtain 
\be
\frac{a}{f}\, F_{\mu\nu}\tilde F^{\mu\nu} \xrightarrow[]{\text{int. by parts}} -2\frac{\partial_\mu a}{f}\, A_{\nu}\tilde F^{\mu\nu} \ ,
\label{operator1Bis}
\ee
and the right-hand side is indeed PQ-invariant, so it does not add up to the PQ shift of the action but still affects the physical amplitudes. Thus, if we integrated by parts a fraction of \eqref{operator1} to write our EFT as follows,
\be
\cL \supset -g^2\frac{D^{\text{PQ}AA}}{16\pi^2f}a F_{\mu\nu}\, \tilde F^{\mu\nu}+g^2\frac{\cC-D^{\text{PQ}AA}}{8\pi^2f}\partial_\mu a\, A_\nu \, \tilde F^{\mu\nu} \ ,
\label{EFTarranged1}
\ee
we would obtain the same matrix elements as if we used \eqref{operator1}, however now we get the expected shift under $U(1)_\text{PQ}$, namely \eqref{anomalousShiftAxion} with $\cC\rightarrow D^{\text{PQ}AA}$, consistently with anomaly matching.

It may be puzzling that we modified an anomalous shift using integration by parts only. This is however perfectly consistent, since, for a constant $\epsilon_\text{PQ}$, \eqref{anomalousShiftAxion} is a total derivative term. The same is true for the gauge variation of the second term in \eqref{EFTarranged1}. Thus, what should be understood as the genuine anomalous variation of the axion EFT is unclear at this stage. Indeed, it requires a further formal step that we discuss in section \ref{section:PQMatching}, where we study anomaly matching when the PQ transformation is made local. There, we also justify the need for the splitting in \eqref{EFTarranged1}. However, for the time being, we simply performed this artificial reshuffling of the EFT to suggest that anomalies may not capture all of the axion-gauge bosons couplings, in particular that they may not be enough to obtain the full axion decay amplitude when such a decay can occur, as was found in \cite{Quevillon:2019zrd}. 

\subsection{UV-IR matching}\label{section:AbelianMatching}

As a first step towards more quantitative statements, we study in this section the matching of processes involving an axion and two gauge fields, comparing UV theories and their IR EFTs. We present general formulae that will be used in the next sections.

We derive the relevant EFT terms from a simple UV toy model, the theory of a charged massive fermion $\psi$ which obtains its mass from a Yukawa coupling to a Higgs field $\phi$:
\be
\cL_\psi = i\overline{\psi}\gamma^\mu\left(\partial_\mu-ig[\alpha-\beta\gamma_5]A_{\mu}\right)\psi-y(\overline{\psi_{L}}\psi_{R}\phi+h.c.) \ ,
\label{lagHeavyFermion}
\ee
where $\psi_{R/L}=\frac{1\pm\gamma_5}{2}\psi$ have charges $q_{R/L}=\alpha\mp\beta$ with respect to the gauge field $A_{\mu}$, and we choose the Yukawa coupling $y$ to be real. \eqref{lagHeavyFermion} also has a global PQ symmetry, under which $\psi_{R/L}$ have charges $q^\text{PQ}_{R/L}=\alpha_\text{PQ}\mp\beta_\text{PQ}$ and $\phi$ has charge $q^\text{PQ}_\phi=q^\text{PQ}_L-q^\text{PQ}_R=2\beta_\text{PQ}$. With those charges, the mixed anomaly coefficient between one PQ current and two gauge currents reads
\be
D^{\text{PQ}AA}=2([\alpha^2+\beta^2]\beta_\text{PQ}+2\alpha\beta\alpha_\text{PQ}) \ .
\label{PQAAanomalyCoefficient}
\ee
When $\phi$ gets a vev $\langle\phi\rangle=\frac{f}{\sqrt{2}}$, which we also choose to be real (both $y$ and $f$ can be made real by constant phase rotations of the fermions and the Higgs field), $\psi$ acquires a mass $m_\psi=\frac{yf}{\sqrt{2}}$. By integrating this fermion out\footnote{To have our computations under control, we focus on perturbative theories, and in parameter spaces which consistently allow to integrate out the fermions. Namely, we choose Yukawa couplings for the fermions so that the theory is perturbative (i.e. we take $y\lesssim 4\pi$ \cite{Manohar:1983md,Luty:1997fk,Cohen:1997rt,Gavela:2016bzc}). Then, the scale $m_\psi\sim yf$ defines the UV cutoff scale of the axion-gauge bosons EFT, and each process we consider involves energies much below this scale. For the EFT at this scale to still be made out of axions and gauge bosons, we choose small enough values for the gauge couplings (and for any symmetry-breaking parameter which gives axions a mass). If we want to keep the radial excitations of the Higgs fields in the EFT, we can also choose their quartic couplings accordingly. Perturbative unitarity constraints, which forbid certain decoupling limits \cite{Preskill:1990fr,DiLuzio:2016sur,Falkowski:2019tft,Craig:2019zkf}, can also be enforced without affecting our discussion.}, we get the low energy EFT of axions and gauge fields \cite{DHoker:1984izu,DHoker:1984mif} which we can match to our previous EFT discussion.

Before we proceed, let us pause and comment on two aspects of \eqref{lagHeavyFermion}. The first one is that we do not enforce gauge anomaly cancellation at the level of \eqref{lagHeavyFermion} (which would mean $\beta=0$), for several reasons. First, anomaly cancellation would impose some relations between the charges which may hinder or wrongly suggest a rationale for extracting axion couplings from anomalies. Second, we could be describing non-dynamical gauge fields, sources for the symmetry currents, so that anomaly cancellation is irrelevant. This will actually be our interpretation of a PQ gauge field in section \ref{section:PQMatching}. Finally, there could be additional light fermions in the theory, or other heavy fermions, or even a fundamental Green-Schwarz scalar \cite{Green:1984sg,Faddeev:1986pc,Krasnikov:1985bn,Babelon:1986sv,Harada:1986wb,Bonnefoy:2018hdo}, so that anomaly cancellation is ensured. Such additional anomalous modes would not affect our computations, which are first order in a perturbative regime, so that we do not have to be definite about their presence or their properties. 

The second comment is related: we leave open the possibility that there are several Higgs fields, so that the one in \eqref{lagHeavyFermion} is not the only one participating in the spontaneous breaking of the gauge and PQ symmetries. This means that the phase of $\phi$ does not need to be fully absorbed by the gauge field, or fully aligned with a physical axion; we only assume that it is the only phase which couples to $\psi$ (more general options are not needed for our purposes). Identifying the massive eigenstates, removing kinetic mixing, or even fixing the gauge in the bosonic Lagrangian is left for after we integrate out the fermion $\psi$. 

These remarks are important for the reader that may be puzzled by two aspects of \eqref{lagHeavyFermion}: first, that there is a gauge anomaly and second, that there is no axion since there is one unbroken combination of the PQ and the gauge symmetries\footnote{Namely, $\beta_\text{PQ}U(1)_A-\beta U(1)_\text{PQ}$ under which $\phi$ is neutral.}. These two puzzles are solved by introducing in the UV enough fermions and enough scalars so that there is no physical gauge anomaly -- but only mixed PQ anomalies -- and that both the gauge and the PQ symmetries are spontaneously broken, as would happen in a full UV model. Starting from \eqref{lagHeavyFermion}, this is, for instance, easily done by adding another fermion $\psi'$ and another scalar $\phi'$ (of vev $v'$) coupled as in \eqref{lagHeavyFermion},
\be
\cL_{\psi'} = i\overline{\psi'}\gamma^\mu\left(\partial_\mu-ig[\alpha'-\beta'\gamma_5]A_{\mu}\right)\psi'-y'(\overline{\psi'_{L}}\psi'_{R}\phi'+h.c.) \ .
\label{lagHeavyFermionComplete}
\ee
Then, the physical gauge anomaly is cancelled by choosing $3\alpha^2\beta+\beta^3+3\alpha'^2\beta'+\beta'^3=0$, both the gauge and the PQ symmetries are broken, the axion $a$ is the gauge-invariant combination of the phases $\frac{\theta}{v},\frac{\theta'}{v'}$ of $\phi,\phi'$ respectively,
\be
a\propto \beta'\frac{\theta}{v}-\beta\frac{\theta'}{v'} \ .
\label{physicalAxionExample}
\ee
Nevertheless, in this simple model, the 
{bosonic effective interactions generated at one-loop by $\psi$}
 are insensitive to the presence of the other fermion $\psi'$ or the other scalar $\phi'$ {(indeed, there is no tree-level mixing between $\psi$ and $\psi'$, and no tree-level coupling between $\phi'$ and $\psi$ or $\phi$ and $\psi'$, so that diagrams with one loop of fermions, which give the leading EFT interactions, can be computed in the primed and unprimed sectors independently)}, so that we can completely forget about them when integrating $\psi$ out. In particular, its contributions to the EFT below its mass can be expressed in terms of $\theta$ only. Consequently, in the following part of this section, we consider only \eqref{lagHeavyFermion}. On the other hand, adding \eqref{lagHeavyFermionComplete} is interesting to show the mismatch between anomalies and axion couplings in the simplest consistent UV setting.

Let us now derive the relevant couplings of the low-energy EFT below the fermion mass. Anomalies are extracted by computing triangle loop diagrams of heavy fermions. Axion-gauge bosons couplings also arise via similar diagrams. Therefore, we compute the (relevant) leading order, dimension 5 terms which arise in the EFT below $m_\psi$. Writing down $\phi=\frac{f}{\sqrt{2}}e^{i\frac{\theta}{f}}$, the couplings between the axion $\theta$ and the gauge field $A_\mu$ is (see appendix \ref{appendix:detailsIntegration} for details)
\be
\cL_\text{EFT}\supset
-g^2\frac{3 \alpha^2 + \beta^2}{48\pi^2}\frac{\theta}{f}F\tilde{F} \ .
\label{EFT:afterFermionNoPQ}
\ee
To connect with section \ref{physicalEFTsection:abelianIntro}, it is useful to rewrite \eqref{EFT:afterFermionNoPQ} under the form \eqref{EFTarranged1}. First, we normalize the PQ charges and fix $\beta_\text{PQ}=\frac{1}{2}$ so that $\delta_\text{PQ}\theta=2\epsilon_\text{PQ}\beta_\text{PQ}f=\epsilon_\text{PQ}f$ as in section \ref{physicalEFTsection:abelianIntro}, and we identify the two coefficients in \eqref{EFTarranged1}:
\be
D^{\text{PQ}AA}=2\left(\frac{\alpha^2+\beta^2}{2}+2\alpha\beta\alpha_\text{PQ}\right) \ , \quad \cC-D^{\text{PQ}AA}=-2\beta\left(2\alpha\alpha_\text{PQ}+\frac{\beta}{3}\right) \ .
\label{splittingAxionEFT}
\ee
With this expression, we immediately observe the following thing: if the spectrum is vector-like with respect to $A$, i.e. if $\beta=0$, the second term vanishes. Consistently, the axion term \eqref{EFT:afterFermionNoPQ} shifts as
\be
\delta_\text{PQ}\left(-g^2\frac{\alpha^2}{16\pi^2}\frac{\theta}{v}F\tilde{F}\right)=-g^2\epsilon_\text{PQ}\frac{\alpha^2}{16\pi^2} F\tilde{F} \ ,
\label{axionTermPQShift}
\ee
where we recognize in this expression $D^{\text{PQ}AA}=2\alpha^2\beta_\text{PQ}=\alpha^2$, the mixed anomaly coefficient between one PQ current and two gauge currents. Thus, the full axion term reproduces the anomaly, as naively expected. In particular, this holds for a massless gauge field, in which case the consistency of the couplings demand that $\beta=0$ (otherwise the vev of $\phi$ breaks the gauge symmetry). On the other hand, this changes for a chiral gauge field, namely one with $\beta\neq 0$. In this case, the shift of the axion term in \eqref{EFT:afterFermionNoPQ} does not generically reproduce $D^{\text{PQ}AA}$. Since the vev of $\phi$ breaks any gauge symmetry such that $q_\phi=q_L-q_R=2\beta\neq 0$, this chiral gauge field becomes massive together with the fermion. We will show in section \ref{section:PQMatching} that those observations are not specific to our model and hold whatever UV model we consider.

One important consequence of this computation is that we can read off the choice of PQ symmetry whose anomaly coefficient reproduces the axion matrix elements. Indeed, in theories with a chiral gauge field, vector-like symmetries can be anomalous, so that they represent an irreducible ambiguity when defining the PQ charges of the fermions and it makes sense to talk about a choice in the PQ symmetry (see \cite{Quevillon:2019zrd,Quevillon:2020hmx} for the case of lepton or baryon numbers). For our case, the $\psi$-number symmetry $\psi\rightarrow e^{i\epsilon}\psi$ is anomalous if $A$ is chiral, consistently with the fact that the vector component of our PQ symmetry, $\alpha_\text{PQ}$, enters in $D^{\text{PQ}AA}$. We can use this freedom to make $\cC-D^{\text{PQ}AA}$ vanish: indeed, from \eqref{splittingAxionEFT}, we see that choosing $6\alpha_\text{PQ}\alpha+\beta=0$ is enough. This ensures that the shift of the axion term fully reproduces $D^{\text{PQ}AA}$. When the normalization of the PQ charges is not specified and $\beta_\text{PQ}$ left generic, this becomes $3\alpha_\text{PQ}\alpha+\beta\beta_\text{PQ}=0$, as seen from the PQ variation of \eqref{EFT:afterFermionNoPQ}:
\begin{eqnarray}
&&\delta_\text{PQ}\left(-g^2\frac{3 \alpha^2 + \beta^2}{48\pi^2}\frac{\theta}{f}F\tilde{F}\right)=-g^2\epsilon_\text{PQ}\frac{(3 \alpha^2 + \beta^2)q^\text{PQ}_\phi}{3}\frac{F\tilde{F}}{16\pi^2}=-g^2\epsilon_\text{PQ}\frac{D^{\text{PQ}AA}}{16\pi^2}F\tilde{F}\nonumber\\ 
&&\text{ when } 3\alpha_\text{PQ}\alpha+\beta_\text{PQ}\beta=0 \ .
\label{prescriptionFormula}
\end{eqnarray}
This prescription for defining the PQ symmetry is general and applicable to any perturbative matching between an anomalous UV sector and an axion EFT. {At this stage, we should pause and wonder if the interplay with gauge anomalies complicates the discussion. Indeed, the fermion in Eq.~\eqref{lagHeavyFermion} carries a gauge anomaly, and the axion term in Eq.~\eqref{EFT:afterFermionNoPQ} generates both a Wess-Zumino-Witten term and a physical axion coupling, since $\theta$ is generically a combination of the longitudinal component of $A_\mu$ and of the physical axion $a$. Consistently, $\cC=\frac{D^{AAA}}{6\beta}$, where $D^{AAA}=2(3\alpha^2+\beta)\beta$ is the $U(1)_A^3$ anomaly coefficient, therefore $\cC$ also corresponds to this UV anomaly coefficient. To clarify the picture, it is useful to consider the addition of \eqref{lagHeavyFermionComplete}, so that one integrates out an anomaly free set of fermions. Summing the different contributions to Eq.~\eqref{EFT:afterFermionNoPQ}, we find the following EFT term
\beq
\cL_\text{EFT}\supset-\frac{1}{48\pi^2}\(\frac{(3\alpha^2+\beta^2)\beta'}{v^2}-\frac{(3\alpha'{}^2+\beta'{}^2)\beta}{v'{}^2}\)\frac{a}{V}F\tilde F \ , \quad \text{with } V\equiv\sqrt{\frac{\beta'{}^2}{v^2}+\frac{\beta^2}{v'{}^2}} \ ,
\label{EFTwithNoGaugeAnomaly}
\eeq
where the axion in Eq.~\eqref{physicalAxionExample} has been canonically normalized. It is straightforward to check that $D^{AAA}$ can vanish (when $3\alpha^2\beta+\beta^3+3\alpha'^2\beta'+\beta'^3=0$, as said earlier) while the axion coupling $\cC$ remains generically non-zero, so that the latter is not anymore proportional to the former in the full EFT. In addition, one can compare $\cC$ with the UV PQ anomaly coefficient,
\be
D^{\text{PQ}AA}=2\(\[\alpha^2+\beta^2\]\beta_\text{PQ}+2\alpha\beta\alpha_\text{PQ}+\[\alpha'{}^2+\beta'{}^2\]\beta'_\text{PQ}+2\alpha'\beta'\alpha'_\text{PQ}\) \ .
\ee
Since $\cC$ does not depend on the PQ charges $\alpha^{(\prime)}_\text{PQ},\beta^{(\prime)}_\text{PQ}$, while the ambiguity associated to vector-like contributions to $\alpha^{(\prime)}_\text{PQ}$ modifies $D^{\text{PQ}AA}$, both coefficients cannot be matched for all PQ charge assignments. Nevertheless, one can choose suitably the PQ charges to ensure $\cC=D^{\text{PQ}AA}$, generalizing Eq.~\eqref{prescriptionFormula}.}

We present examples of this matching prescription in appendix \ref{PQMatchingExampleAppendix}, and it can be generalized to multiple abelian gauge fields and heavy fermions, as explained in appendix \ref{appendix:prescription}. {Let us report here one important output of this analysis: when several gauge fields are present, an important property of the prescription which generalizes Eq.~\eqref{prescriptionFormula} is that there does not always exist a choice of UV PQ charges such that the couplings of the axion to all pairs of gauge fields match all mixed-PQ anomaly coefficients simultaneously. This can be anticipated by looking at Eq.~\eqref{prescriptionFormula}: the appropriate choice of PQ charges depends on the gauge charges, therefore, when there are several of the latter, the requirement from \eqref{prescriptionFormula} for different pairs of gauge fields can be contradictory. The results of appendix \ref{PQMatchingExampleAppendix} provide such an example, and so does the model of Eqs.~\eqref{lagHeavyFermion}-\eqref{lagHeavyFermionComplete} when the two unbroken vector-like symmetries are gauged.} We also comment further on 
the prescription in section \ref{section:GaugedAbelianMatching}.

\subsection{PQ anomaly matching}\label{section:PQMatching}

Now that we have established the generic mismatch between the axion EFT couplings and the UV anomaly coefficients, we discuss in details how the PQ anomaly matching is ensured when matching between the UV and the EFT\footnote{Indeed, although we showed that the UV PQ anomaly coefficient matches the axion EFT coupling for {\it some} PQ charge assignments only, {\it all} PQ anomalies must match, irrespective of the charge assignment.}. This will allow us to justify the splitting in \eqref{EFTarranged1} and to rephrase our previous results.

Let us start again by an EFT analysis, continuing section \ref{physicalEFTsection:abelianIntro}. As we said above, the PQ anomalous shift of the EFT in \eqref{anomalousShiftAxion} is a total derivative for constant $\epsilon_\text{PQ}$, consistently with the fact that \eqref{operator1} and the PQ-invariant \eqref{operator1Bis} are equal up to a boundary term, since they are equivalent up to integration by parts. However, in perturbation theory we are only sensitive to ``bulk'' physics, namely we cannot tell apart an operator from its counterparts obtained after integration by parts. To fully grasp the difference between the two terms in \eqref{EFTarranged1}, we would like to discuss objects which are captured by perturbation theory, and by the rules of the perturbative matching between EFTs and UV theories. In particular, we would only allow ourselves to discuss shifts with non-constant $\epsilon_\text{PQ}$ in \eqref{anomalousShiftAxion}, that are usually considered when discussing gauge theories. Consequently, we couple the UV theory to an auxiliary gauge field $A_\mu^\text{PQ}$, minimally coupled with a gauge coupling $g_\text{PQ}$ to all PQ-charged fields. This field is not associated to a physical propagating particle, it should only be thought of as a classical source that we use as a device to keep track of the anomalous shifts of the action, both in the UV and in the IR. In particular, it is not integrated over in the path integral and is only an argument of the latter, which we can simply put to zero when discussing the physical EFT. Nevertheless, with $A_\mu^\text{PQ}$ the UV action now has a classical ``fake''\footnote{Henceforth, we use the words ``fake'' and ``physical'' to refer to fields which are understood as external classical sources such as $A^\text{PQ}$, and to fields associated to physical particles such as $A$, respectively.} PQ gauge invariance, $\epsilon_\text{PQ}$ can be upgraded to an arbitrary function of spacetime so that $\delta_\text{PQ}A_\mu^\text{PQ}=\frac{1}{g_\text{PQ}}\partial_\mu\epsilon_\text{PQ}$, and \eqref{anomalousShiftAxion} no longer is a total derivative but becomes a PQ anomalous gauge transformation. 

We can now use the properties of the ``fake'' gauge theory to constrain the EFT couplings. In particular, when we match a UV theory to an IR EFT and for the kind of one-loop calculations which are relevant for anomalies, $A_\mu^\text{PQ}$ is equivalent to a physical gauge field from the point of view of the particles that we integrate out. Thus, anomaly matching must hold between the UV and the IR for consistency of the fake gauge theory\footnote{To make this new gauge theory consistent at the quantum level, we can add new massless charged fermions to the theory so as to cancel any (mixed) gauge anomaly. They are charged under the fake PQ gauge field $A_\mu^\text{PQ}$ but also under the physical one $A$, so they change the IR physics, however in perturbative theories they do not affect the one-loop calculations that we focus on and which are relevant for anomalies, provided they do not couple to the axion or the heavy particles. Thus, constraints on the EFT derived when those light fermions are present also hold without them. On the other hand, when they are present, their contribution to the anomalies is identical in the UV and in the IR theories, so that the EFT must retrieve the anomalous contributions of the heavy particles for the consistency of the IR gauge theory.}. Also, the mixed anomaly coefficient $D^{\text{PQ}AA}$ of the UV fields is the same, irrespective of whether the PQ symmetry is gauged or global. We conclude from this that the {\it gauge} PQ shift of the EFT {\it must} be given by \eqref{anomalousShiftAxion} with $\cC\rightarrow D^{\text{PQ}AA}$. Since $\cC \neq D^{\text{PQ}AA}$ in some cases, as was discussed previously and explicitly represented in \eqref{EFTarranged1}, the only consistent possibility is that, in those cases, there are additional anomalous contributions to the PQ shift of the effective action, beyond \eqref{anomalousShiftAxion}. Notice that ambiguities due to integration by parts are not relevant here, since we consider local PQ transformations: both terms in \eqref{EFTarranged1} shift so as to reproduce precisely \eqref{anomalousShiftAxion} in the bulk, with a shift coefficient $\cC$ and not $D^{\text{PQ}AA}$. Thus, we need something beyond axion terms so that the full PQ anomalous shift in the EFT is \eqref{anomalousShiftAxion} with $\cC\rightarrow D^{\text{PQ}AA}$, otherwise we are back to the contradictory requirement that $\cC= D^{\text{PQ}AA}$.

In other words, we are left with the question: how can we make the second term in \eqref{EFTarranged1} PQ- and gauge-invariant, so that only the first term contributes to the PQ anomaly in the EFT? Ensuring the PQ invariance is easy now that we introduced the PQ gauge field, we simply need to turn the derivative that acts on the axion field into a covariant one:
\be
\frac{\partial_\mu a}{f}\, A_{\nu}\tilde F^{\mu\nu} \xrightarrow[]{} \left(\frac{\partial_\mu a}{f}-g_\text{PQ} A_\mu^\text{PQ}\right)\, A_{\nu}\tilde F^{\mu\nu} \ .
\label{operator1Ter}
\ee
On the other hand, this modification breaks the gauge-invariance, and we should further modify \eqref{operator1Ter} to restore it. This can be achieved, but only when $A$ is massive, since then one can use the longitudinal polarisation of the gauge field $A_\mu$ \cite{Callan:1984sa,Preskill:1990fr,Craig:2019zkf}. Starting from a unitary gauge, this is easily seen by performing a Stueckelberg trick, namely reinstating via a gauge transformation the Goldstone boson $\theta_A$ which makes up for the longitudinal component of the gauge field:
\be
\cL_\text{EFT} \supset -\frac{m_A^2}{2}A_\mu^2 \xrightarrow[]{A_\mu\rightarrow A_\mu-\frac{\partial_\mu\theta_A}{gm_A}} -\frac{1}{2g^2}(\partial_\mu\theta_A-gm_AA_\mu)^2 \ .
\label{Stueckelberg}
\ee
$\theta_A$ is charged under $U(1)_A$, $\delta_A\theta_A=\epsilon_Am_A$, and the mass term of $A$ is understood as coming from the kinetic term of $\theta_A$, as familiar from the Brout-Englert-Higgs mechanism. By applying the same transformation to the right-hand side of \eqref{operator1Ter}, it becomes
\be
-\frac{1}{g}\left(\frac{\partial_\mu a}{f}-g_\text{PQ} A_\mu^\text{PQ}\right)\, \left(\frac{\partial_\nu\theta_A}{m_A}-gA_\nu\right)\tilde F^{\mu\nu} \ ,
\label{operator1Quator}
\ee
which is now fully gauge and PQ invariant. The existence of \eqref{operator1Quator} makes it clear that anomalies may not capture all of the axion-gauge bosons couplings, in particular that they may not be enough to obtain the full axion decay amplitude when such a decay can occur, as was found in \cite{Quevillon:2019zrd}. Indeed, \eqref{operator1Quator} contributes to the physical amplitudes but not to the anomalies. When imposing $A_\mu^\text{PQ}=0$ to recover the physical EFT, \eqref{operator1Quator} degenerates to the right-hand side of \eqref{operator1Bis}\footnote{The terms that depend on $\theta_A$ vanish by virtue of the antisymmetry of $\tilde F^{\mu\nu}$ and the Bianchi identity:
\be
-\frac{1}{g}\frac{\partial_\mu a}{f}\bigg(\frac{\partial_\nu\theta_A}{m_A}-gA_\nu\bigg)\, \tilde F^{\mu\nu} \xrightarrow[]{\text{int. by parts}} \frac{1}{g}\frac{a}{f} \bigg(\frac{\partial_{[\mu}\partial_{\nu]}\theta_A}{m_A}-g\frac{F_{\mu\nu}}{2}\bigg)\, \tilde F^{\mu\nu}=-\frac{1}{2g}a F_{\mu\nu}\, \tilde F^{\mu\nu} \ .
\ee
}: as announced, this shows that the latter has a very different anomalous structure than the left-hand side. Indeed, it does not shift, neither under the PQ symmetry nor under the gauge symmetry, consistent with the fact that we obtained it from \eqref{operator1Quator}. The two terms in \eqref{operator1Bis} differ by a boundary term, whose PQ variation precisely cancels that of the left-hand side of \eqref{operator1Bis} and makes the right-hand side PQ invariant. Since we cannot constrain boundary terms when building the EFT, using $A_\mu^\text{PQ}$ is a useful trick to sort out what terms carry the anomalies, as we will explicitly see by matching to a UV theory in a few lines.

Thus, we like to think of the EFT of an axion and a massive gauge field as being split as follows, 
\be
\cL \supset -g^2\frac{D^{\text{PQ}AA}}{16\pi^2f}a F_{\mu\nu}\, \tilde F^{\mu\nu}-g\frac{\cC-D^{\text{PQ}AA}}{8\pi^2f}\partial_\mu a\, \left(\frac{\partial_\nu\theta_A}{m_A}-gA_\nu\right) \, \tilde F^{\mu\nu} \ ,
\label{EFTarranged}
\ee
which is completely equivalent to \eqref{operator1} at the level of scattering amplitudes, but which makes the (non-)anomalous nature of each term explicit as argued above.

Eventually, coupling the theory to an external PQ gauge field and ensuring that the PQ symmetry is the only one broken by the anomalies, we have confirmed some of our previous conclusions, namely
\begin{itemize}
\item for a massless gauge field $A$, there cannot be any non-anomalous EFT term beyond \eqref{operator1}, so this operator generates both the full anomalous shift of the UV theory, as well as the full (dimension-five) axion-gauge field amplitudes. Anomaly matching is thus expected to capture its coefficient, in particular anomalies always capture the dimension-five coupling to photons or gluons, 
\item for a massive gauge field, the gauge and PQ invariant operator in \eqref{operator1Quator} is available, and generates new contributions to the axion-gauge field amplitudes beyond that of \eqref{operator1}. Although it is equivalent to \eqref{operator1} from the point of view of scattering amplitudes, it is not anomalous. Nevertheless, from the EFT, we cannot understand its precise coefficient.
\end{itemize}

Let us end this section by a remark which connects to our explicit computation in section \ref{section:AbelianMatching}. We saw there that $\beta=0$ is enough to ensure that the coefficient of the axion term is simply given by the anomaly. Therefore, the fact that the gauge field is massive and has a longitudinal part does not suffice to allow for $\cC\neq D^{\text{PQ}AA}$, it also has to be chiral with respect to some heavy fermion. Note that this last condition implies that the gauge field has a mass, otherwise the fermion cannot have a mass either\footnote{In the vector-like case, the mass of the vector can be sent to zero by adjusting a parameter such as a Higgs vev without affecting the one-loop EFT below the heavy fermion mass, so that the results must be the same as for a genuinely massless vector.} and there is no notion of integrating it out. 

\subsection{UV-IR matching with a gauged PQ symmetry}\label{section:GaugedAbelianMatching}

Now, we perform a one-loop matching in a theory where the auxiliary PQ gauge field is present, in order to support the conclusions of the previous section. In particular, this allows to confirm that PQ anomaly matching does hold when matching the UV and the EFT. For that, since we already computed the EFT terms which correspond to the restriction of \eqref{operator1Quator} to terms that involve one pseudoscalar, and since those which involve two trivially vanish after integration by parts, we only need to compute the restriction of \eqref{operator1Quator} to gauge fields only: this leads to another anomalous term in the EFT, the GCS term \cite{Anastasopoulos:2006cz}
,
\be
g_\text{PQ}g^2A_\mu^\text{PQ}\, A_\nu\, \tilde F^{\mu\nu} \ .
\label{operator2}
\ee
We included three gauge couplings in \eqref{operator2} to match with the outcome of the UV computation to be discussed in a few lines. This term generates anomalous shifts under both the PQ and the gauge symmetries,
\begin{eqnarray}
&&\delta_\text{PQ}\left(g_\text{PQ}g^2A_\mu^\text{PQ}\, A_\nu\, \tilde F^{\mu\nu}\right) = g^2\partial_\mu\epsilon_\text{PQ}\, A_\nu\, \tilde F^{\mu\nu} \xrightarrow[]{\text{int. by parts}} -g^2\frac{\epsilon_\text{PQ}}{2}F_{\mu\nu}\, \tilde F^{\mu\nu} \ ,\nonumber\\
&&\delta_\text{A}\left(g_\text{PQ}g^2A_\mu^\text{PQ}\, A_\nu\, \tilde F^{\mu\nu}\right) \xrightarrow[]{\text{int. by parts}}  gg_\text{PQ}\frac{\epsilon_A}{2}F_{\mu\nu}^\text{PQ}\, \tilde F^{\mu\nu} \ ,
\label{gaugeVariationsGCS}
\end{eqnarray}
where we made use of the Bianchi identity. In particular, this GCS term is relevant for PQ anomaly matching: its PQ (and gauge) transformations are precisely what make \eqref{operator1Quator} gauge-invariant. Notice that it could only be written after we introduced $A_\mu^\text{PQ}$.

To obtain the GCS term by a direct computation, we extend the model of \eqref{lagHeavyFermion} to incorporate the minimal coupling to $A^\text{PQ}_{\mu}$:
\be
\partial_\mu-ig[\alpha-\beta\gamma_5]A_{\mu}\longrightarrow \partial_\mu-ig[\alpha-\beta\gamma_5]A_{\mu}-ig_\text{PQ}[\alpha_\text{PQ}-\beta_\text{PQ}\gamma_5]A^\text{PQ}_{\mu} \ ,
\ee
and we define again the charges $q_{R/L}^i=\alpha_i\mp\beta_i$ with respect to the gauge field $A^i\in\{A_{\mu},A^\text{PQ}_{\mu}\}$. In the one-loop dimension 5 EFT below the fermion mass, we find GCS terms in addition to the axion term. There are several kinds of GCS terms which are presented in appendix \ref{appendix:detailsIntegration}, however we only need the following EFT terms for our discussion,
\be
\cL_\text{EFT}\supset-g^2\frac{3 \alpha^2 + \beta^2}{48\pi^2}\frac{\theta}{f}F\tilde{F} +g_\text{PQ}g^2\frac{\left(\alpha_\text{PQ}\alpha+\frac{1}{3}\beta_\text{PQ}\beta\right)\beta}{2\pi^2}A_{\text{PQ},\mu}A_{\nu}\tilde{F}^{\mu\nu}\ .
\label{EFT:afterFermion}
\ee
It is easy to check that the PQ variation of \eqref{EFT:afterFermion} reproduces the UV shift, consistently with anomaly matching:
\be
\delta_\text{PQ}\cL_\text{EFT}=-g^2\left(2\frac{3 \alpha^2 + \beta^2}{3}\beta_\text{PQ}+4\left(\alpha_\text{PQ}\alpha+\frac{1}{3}\beta_\text{PQ}\beta\right)\beta\right)\frac{F\tilde{F} }{16\pi^2}=-g^2\frac{D^{\text{PQ}AA}}{16\pi^2}F\tilde{F}  \ .
\ee
One remark is in order at this point: one sees from \eqref{EFT:afterFermion} that no GCS term is generated for a vector-like fermion, for which $\beta=0$. In particular, its anomalous variation under the physical gauge transformation trivially cancels. This is not a surprise, since our computation has been carried out in such a way that this anomalous variation cancels\footnote{Indeed, in UV theories, one always has the freedom to choose which symmetry current carries an anomaly (see e.g. the sections on anomalies of classic textbooks such as \cite{Pokorski:1987ed,Weinberg:1996kr,Zee:2003mt}, or \cite{Kabat:2019tbr}). This can be understood precisely as the freedom to add counterterms such as \eqref{operator2}, related to momentum shift ambiguities in loop diagrams, see appendix \ref{appendix:detailsIntegration} for examples. In particular, in a theory free of gauge anomalies, one can (and should) always choose the regulator and the counterterms such that the effective action does not shift under a gauge transformation. It is for instance done when computing the axial anomaly that determines the pion decay rate to two photons at leading order, and it is a consistency requirement, namely that our theory must be such that massless vectors couple to exactly conserved currents. The normalization of the PQ shift in \eqref{anomalousShiftAxion} depends on this choice.}. This justifies why we demanded that the operator \eqref{operator1Ter} be upgraded into a gauge-invariant one while there were (fake) gauge anomalies in the first place. For a massless gauge field, there does not exist any counterterm that can balance the gauge transformation of the GCS term in \eqref{gaugeVariationsGCS}, so that the latter cannot be included in the EFT and the axion term \eqref{operator1} must yield the full anomalous shifts, consistently with what we found previously when $\beta=0$ (which is compatible with a vanishing mass for the gauge field). This is not true for a massive gauge field (in particular, for a chiral gauge field), since we can now write 
\be
\frac{\theta_A}{m_A}F_{\mu\nu}\tilde F_\text{PQ}^{\mu\nu} \ ,
\label{thetaATerm}
\ee
whose gauge variation can cancel that of the GCS term. Indeed, using the formulae in appendix \ref{appendix:detailsIntegration}, one can check that in our toy model the EFT contains the term
\be
\cL_\text{EFT}\supset -g^2\frac{3 \alpha\alpha_\text{PQ} + \beta\beta_\text{PQ}^2}{24\pi^2}\frac{\theta}{f}F\tilde F_\text{PQ} \ ,
\ee
whose physical gauge variation cancels that of the GCS term in \eqref{EFT:afterFermion}. \eqref{thetaATerm} is the third EFT term that adds up to the axion and GCS terms to build up \eqref{operator1Quator}. Consequently, it is possible to rearrange the EFT so that it looks like \eqref{operator1}+\eqref{operator1Quator}. This is straightforwardly done from our explicit formulae, but it is also simply understood from a field redefinition of the UV theory, as we explain in appendix \ref{CCWZsection}. \eqref{thetaATerm} also has a non-trivial PQ variation, but we ignore it since it concerns a $U(1)_\text{PQ}^2\times U(1)_A$ anomaly which is irrelevant in the physical theory, indeed it vanishes when $A^\text{PQ}_\mu$ is put to zero.

Using the notion of GCS terms, we can also recover one of the results of section \ref{section:AbelianMatching}, namely the fact that there exists a prescription to define a non-generic PQ symmetry that matches $D^{\text{PQ}AA}$ with $\cC$. We can identify this prescription from the PQ variation of the GCS term in \eqref{EFT:afterFermion}: indeed, wee see that choosing $3\alpha_\text{PQ}\alpha+\beta_\text{PQ}\beta=0$ makes the PQ variation of the GCS term vanish,
\be
\delta_\text{PQ}\left(g_\text{PQ}g^2\frac{\left(\alpha_\text{PQ}\alpha+\frac{1}{3}\beta_\text{PQ}\beta\right)\beta}{2\pi^2}A_{\text{PQ},\mu}A_{\nu}\tilde{F}^{\mu\nu}\right)=-g^2\frac{\left(3\alpha_\text{PQ}\alpha+\beta_\text{PQ}\beta\right)\beta}{12\pi^2}\epsilon_{\text{PQ}}F_{\mu\nu}\tilde{F}^{\mu\nu}=0 \ ,
\label{prescriptionGCS}
\ee
where we integrated by parts in the first equality. Consequently, the axion term must fully reproduce $D^{\text{PQ}AA}$, as we already saw. However, we should mention two subtleties related to this prescription. One of them, which we also encounter in appendix \ref{PQMatchingExampleAppendix}, is that constraints such as \eqref{prescriptionFormula} may be too restrictive when applied independently on each massive fermion, for instance when the UV charges $q_\text{PQ}$ are constrained so that $U(1)_\text{PQ}$ commutes with non-abelian gauge symmetries. In this case, one could as well apply \eqref{prescriptionFormula} at the level of a subset of the heavy fermions $\{\psi\}$, $\sum_{\psi \, \in \text{ subset } \{\psi\}}(3\alpha_\text{PQ}^{(\psi)}\alpha^{(\psi)}+\beta_\text{PQ}^{(\psi)}\beta^{(\psi)})=0$, for instance at the level of a complete gauge representation. This is sufficient to ensure that the axion couplings in the EFT below the mass of all the fermions in the subset reproduce the PQ anomalies. This is easily understood using \eqref{prescriptionGCS}, since the contributions of all the fermions add up to the GCS term. Another subtlety arises when $\theta$ is not aligned with a physical axion and must be rotated together with the other pseudoscalars, as for instance in \eqref{physicalAxionExample}. Then, in principle, the condition in \eqref{prescriptionFormula} ensures that the combined PQ shift of all axion terms reproduce the anomaly, but not that the shift of the physical axion term alone suffices. One must then also ensure that the other axions, such as longitudinal components of gauge fields, are PQ neutral. This can usually be ensured by a choice of PQ symmetry, see the example in section \ref{sec:DFSZUV}.

Let us end this section with a remark related to the anomalous shifts of the PQ gauge field. By our direct computation, we reproduced the structure of the GCS term \eqref{operator2}, including the three gauge couplings. This implies that the PQ anomalous variation of the GCS term does not depend on $g_\text{PQ}$ and does not decouple if we remove the interactions of $A_\text{PQ}$ by sending $g_\text{PQ}\rightarrow 0$, so that even in a theory where $A_\text{PQ}$ does not interact, the axion term is still not captured by the anomaly. This additional, non-decoupling shift cannot be easily understood in the EFT unless one introduces $A_\text{PQ}$.

\section{Axion couplings to gauge fields: the Standard Model case}\label{section:axionSM}

In this section, we continue our line or arguments to include couplings of an axion to the electroweak gauge bosons of the SM.

\subsection{Non-anomalous EFT terms for non-abelian theories}\label{introNonAbelian}

Let us begin by suggesting how the discussion of abelian theories of the previous section can be generalized to non-abelian gauge fields. The starting point is, as usual, the dimension five coupling of an axion $a$ to a non-abelian gauge field $A_\mu=A_\mu^aT^a$, where $T^a$ are the generators of the gauge group and we denote $F$ the gauge field strength. It reads:
\begin{equation}
\cL\supset-g^2 \frac{\cC}{16\pi^2} \frac{a}{f} \Tr(F\tilde F) \, .
\label{operator1NA}
\end{equation}
As for the abelian case, this operator is gauge-invariant and seems to shift under a PQ transformation, which is nothing but a total derivative when the PQ symmetry is global. This can be understood given that one can rewrite \eqref{operator1NA} as follows \cite{DHoker:1984izu}:
\be
\cL\supset g^2 \frac{\cC}{8\pi^2}\frac{\partial_\mu a}{f}\, \text{Tr} \left(A_\nu\, \tilde F^{\mu\nu}+\frac{i}{3}\epsilon^{\mu\nu\rho\sigma}A_\nu A_\rho A_\sigma\right) \ .
\label{operator2NA}
\ee
Nevertheless, both expressions again yield the same PQ transformation when the latter is local. To make \eqref{operator2NA} gauge-invariant, we again need that the gauge field have a longitudinal component, so that we can upgrade \eqref{operator2NA} to
\be
\frac{\partial_\mu a}{f}\, \text{Tr} \left(A_\nu\, \tilde F^{\mu\nu}+...\right)\rightarrow -\frac{1}{g}\left(\frac{\partial_\mu a}{f}-g_\text{PQ}A_\mu^\text{PQ}\right)\, \text{Tr} \left(\left[\partial_\mu\theta_A-gA_\nu\right]\, \tilde F^{\mu\nu}+...\right) \ ,
\ee
where we remain deliberately sketchy about the non-abelian structure and hid several non-linear terms. Indeed, the counting of the gauge Goldstone bosons and the structure of their interactions depend on the precise symmetry breaking pattern, and we do not engage in a precise discussion of Wess-Zumino terms that arise when integrating out a fermion coupled to a non-abelian gauge field. Instead, we leave precise examples, as well as the discussion of their phenomenological implication, to the study of the electroweak gauge bosons of the SM and the non-linear representation of its gauge group discussed in section \ref{sumrulesSection}. In the present section, our aim was simply to insist on the fact that the logic that lead us to define non-anomalous terms for abelian theories extends to the non-abelian case.

\subsection{Non-anomalous EFT terms for the SM and sum rules}\label{sumrulesSection}

Our discussion of abelian theories was mostly about splitting one single operator, the one in \eqref{operator1}, into anomalous and non-anomalous parts. However, the phenomenology of the axion, namely the matrix elements between physical states, are all captured by the only coefficient $\cC$, and there is no difference in the phenomenology predicted by anomalous and non-anomalous terms. This remains true when several abelian gauge fields are present since, for abelian theories, there is one mixed PQ anomaly coefficient associated to each axion coupling to a given pair of gauge fields. On the other hand, when non-abelian symmetries are present, things are different, due to the constraining form of the mixed PQ anomalies with any non-abelian factor. For instance there is no $U(1)_\text{PQ}\times SU(2)_L\times U(1)_Y$ anomaly. What we show in this section is that, unlike the abelian case, there are differences in the non-abelian case between the phenomenological predictions of anomalous and non-anomalous EFT operators.

Let us start by displaying the operators analog to \eqref{operator1} that couple an axion to SM electroweak gauge fields,
\be
\cL\supset-\frac{g^2 C_{WW}}{16\pi^2} \frac{a}{f} W^a \tilde W^a -\frac{g'^2 C_{BB}}{16\pi^2} \frac{a}{f} B \tilde B \, .
\label{SManomalousOperators} 
\ee
We only included two coefficients to respect the $SU(2)_L\times U(1)_Y$ gauge symmetry. This is consistent with the constraining form of the mixed PQ anomalies with the $SU(2)$ non-abelian factor of the SM gauge group. 

If those operators were the whole story, they would induce correlations between the different EFT operators when written in terms of the vector massive eigenstates. Indeed, using the latter, \eqref{SManomalousOperators} becomes
\be
-\frac{16\pi^2}{e^2}\cL\ \supset \ C_{\gamma\gamma}\frac{a}{f}F\tilde F+2\frac{C_{Z\gamma}}{c_Ws_W}\frac{a}{f}F\tilde Z+\frac{C_{ZZ}}{c_W^2s_W^2}\frac{a}{f}Z\tilde Z+\frac{2C_{WW}}{s_W^2}W^+\tilde W^- \ ,
\label{SManomalousOperatorsZphotonW}
\ee
where the coefficients read
 \be
 C_{\gamma\gamma}=C_{WW}+C_{BB}\ ,\quad C_{Z\gamma}=c_W^2C_{WW}-s_W^2C_{BB}\ ,\quad C_{ZZ}=c_W^4C_{WW}+s_W^4C_{BB} \ ,
 \label{constraintRelationCoefficients}
 \ee
and where $c_W,s_W,t_W=\cos(\theta_W),\sin(\theta_W),\tan(\theta_W)$ with $\theta_W$ the Weinberg angle, and we called $F$ and $Z$ the photon and $Z$ boson field strengths, respectively. The four coefficients in \eqref{SManomalousOperatorsZphotonW} being determined by $C_{BB}$ and $C_{WW}$ only, there must be correlations between the processes involving one axion and two electroweak gauge fields. This is different from the abelian case, that would allow independent coefficients for each gauge field pair. Such correlations can take the form of sum rules between EFT coefficients,\footnote{{ The sum-rules 
are understood to hold at energy scales where the $W$ and $Z$ bosons are 
dynamical degrees of freedom.}} 
for instance 
\be
C_{\gamma\gamma}+s_W^{-2}(1-t_W^2)C_{Z\gamma}-\frac{1}{s_W^2c_W^2} C_{ZZ} =0\ , \quad C_{\gamma\gamma}+s_W^{-2}C_{Z\gamma}-(1+t_W^{-2})C_{WW} =0 \ .
\label{sumRuleEFTCoeffs}
\ee
Sum rules can also be written at the level of observable quantities such as partial decay rates of the axion, see section \ref{axionCouplingsPheno}.
 
However, \eqref{SManomalousOperators}  does not capture all the possible processes, since there also exist non-abelian analogs of \eqref{operator1Quator}. Indeed, setting $A^{\text{PQ}}=0$ to focus on physical fields, notice first that we can rewrite \eqref{operator1Quator} as follows,
\be
\frac{i}{2gf}\partial_\mu a D_\nu U U^\dagger \tilde F^{\mu\nu}+h.c. \ ,
\ee
provided we define a unitary matrix $U\equiv e^{i\frac{\theta_A}{m_A}}$ and its covariant derivative $D_\mu U=(\partial_\mu -igA_\mu)U$. This manipulation can be extended to the SM non-abelian gauge fields, namely we can write the following PQ- and gauge-invariant operators \cite{Brivio:2017ije},
\be
\partial_\mu a \, \text{Tr}(TV_\nu) \, \tilde B^{\mu\nu} \ , \quad \partial_\mu a \, \text{Tr}(V_\nu \, \tilde W^{\mu\nu}) \ , \quad \partial_\mu a \, \text{Tr}(TV_\nu)\text{Tr}(T \tilde W^{\mu\nu})\ , 
\label{termsToBeAdded}
\ee
where, using the matrix $U$ containing the longitudinal components $\pi^a$ of the $Z$ and $W$ bosons and its covariant derivative \cite{Feruglio:1992wf,Brivio:2017ije},
\be
U=e^{i\frac{\pi^a}{v}\sigma^a} \ , \quad D_\mu U=\partial_\mu U -ig W_\mu U +ig' B_\mu U\frac{\sigma_3}{2} \ ,
\ee
we defined
\be
V_\mu=D_\mu U U^\dagger \ , \quad T=U\sigma_3 U^\dagger \ .
\ee
The $\sigma^a$ are the Pauli matrices and $v\approx 246$ GeV is the electroweak vev. The terms in \eqref{termsToBeAdded} generate amplitudes involving one axion and two gauge bosons in addition to that of \eqref{SManomalousOperators}, since
\be
\bead
\partial_\mu a \, \text{Tr}(TV_\nu) \, \tilde B^{\mu\nu} &\supset -i\partial_\mu a \, (gW^3_\nu-g'B_\nu) \tilde B^{\mu\nu}\big\vert_\text{lin.} \ , \\
\quad \partial_\mu a \, \text{Tr}(V_\nu \, \tilde W^{\mu\nu}) &\supset -\frac{i}{2}\partial_\mu a \, (gW^a_\nu\tilde W^{a,\mu\nu}-g'B_\nu \tilde W^{3,\mu\nu}) \big\vert_\text{lin.}  \ , \\
\quad \partial_\mu a \, \text{Tr}(TV_\nu)\text{Tr}(T \tilde W^{\mu\nu}) &\supset -\frac{i}{2}\partial_\mu a \, (gW^3_\nu-g'B_\nu )\tilde W^{3,\mu\nu})\big\vert_\text{lin.}\ ,
\eead
\label{firstTermsNonLinear}
\ee
where we used the same letter to refer to the gauge fields and to their field strengths, and where $W^a_{\mu\nu}\vert_\text{lin}=\partial_\mu W^a_\nu-\partial_\nu W^a_\mu$ (the couplings of an axion to two gauge fields are only sensitive to the linear parts of the field strengths, so we ignore non-linear pieces). Up to numerical coefficients, those terms respectively integrate by parts to
\be
-\frac{c_1g'}{16\pi^2f}a (gW^3-g'B)\tilde B\ , \quad -\frac{c_2g}{16\pi^2f}a (gW^a\tilde W^a-g'B\tilde W^3) \ , \quad -\frac{c_3g}{16\pi^2f}a(g W^3-g'B) \tilde W^3 \ ,
\ee
where all the field strengths are understood to be restricted to their linear pieces, as in the rest of this paper. Added up and rewritten using the vector massive eigenstates, they lead to
\be
-\frac{e^2}{16\pi^2}\frac{a}{f}\bigg(\frac{c_1+c_2+c_3}{c_Ws_W}\frac{a}{f}F\tilde Z+\frac{(c_2 + c_3) c_W^2 - c_1 s_W^2}{c_W^2s_W^2}\frac{a}{f}Z\tilde Z+\frac{2c_2}{s_W^2}W^+\tilde W^-\bigg) \ .
\label{newTerms}
\ee
It should be noted that the expression above does not contain any $aF\tilde F$ term, consistently with our previous analysis: anomalous terms capture all the processes involving one axion and two photons. In addition, the new terms increase the parameter space of axion EFTs, for instance they generically violate sum rules such as \eqref{sumRuleEFTCoeffs}: using \eqref{newTerms} we find
\be
C_{\gamma\gamma}+s_W^{-2}(1-t_W^2)C_{Z\gamma}-\frac{1}{s_W^2c_W^2} C_{ZZ}=\frac{c_1 - c_2 - c_3}{2c_W^2s_W^2}\ ,\quad C_{\gamma\gamma}+s_W^{-2}C_{Z\gamma}-(1+t_W^{-2})C_{WW} =\frac{c_1-c_2+c_3}{2s_W^2} \ .
\label{sumRulesViolationEFTCoeffs}
\ee
We will give a precise example of that in section \ref{axionCouplingsPheno}. This shows that, unlike what happens in the abelian case, there are genuine differences between the phenomenological predictions using only \eqref{SManomalousOperators} and those that also include the terms in \eqref{newTerms}. For instance, the use of the non-linear realization of the SM gauge group allowed for the axion couplings to go beyond the $SU(2)$ trace structure in \eqref{SManomalousOperators}, so that the non-anomalous terms cannot simply be integrated by parts to recover the usual phenomenology of the anomalous terms. In particular, the manipulation in section \ref{introNonAbelian} leads to one non-anomalous term within a larger series.

There are two general properties of the non-anomalous terms in \eqref{operator1Quator} and \eqref{termsToBeAdded} that are worth commenting here, namely: when do those terms arise? and with what coefficient do they enter the EFT? 

The answers to these questions are related, so let us start with the first one. In a pure EFT approach, the non-anomalous terms in \eqref{operator1Quator} and \eqref{termsToBeAdded} only make use of the Goldstone polarizations of the massive gauge fields, which exist for both vector-like or chiral vectors, as long as they are massive. However, as we saw in section \ref{section:AbelianMatching}, their presence is intrinsically related to integrating out chiral fermions. As we said, vectors must be massive if there exist chirally charged massive fermions, but on the contrary, the vector mass is irrelevant if the charged particles are all vector-like, in which case the non-anomalous terms are not generated at leading order and the anomalies capture the dimension-5 axion couplings. That means in particular that a violation of the relations \eqref{constraintRelationCoefficients}, for instance, via a violation of the sum-rule \eqref{sumRuleEFTCoeffs} is a smoking gun of the presence of heavy chiral matter. 

This link with chiral matter explains why we chose to represent the non-anomalous axion couplings to SM fields as in \eqref{termsToBeAdded}. Indeed, it is known that a non-linear representation of the SM d.o.f.s (i.e. a HEFT \cite{Feruglio:1992wf}) is necessary when one integrates out matter whose mass comes entirely from Higgs fields \cite{Cohen:2020xca}. Had we ignored this, we could have instead tried to write the operators in the right-hand side of \eqref{firstTermsNonLinear} using the full Higgs doublet of the SM, since those terms can be obtained as the first ones in an expansion using either linear or non-linear realizations of the SM gauge group \cite{Brivio:2017ije}. The latter case corresponds to \eqref{termsToBeAdded}, while in the former case, they arise from
\be
\partial_\mu a \left(H^\dagger \overleftrightarrow D_\nu H\right) \tilde B^{\mu\nu} \ , \quad \partial_\mu a \left(D_\nu H^\dagger \tilde W^{\mu\nu} H-H^\dagger \tilde W^{\mu\nu} D_\nu H\right) \ , \quad
\partial_\mu a \left(H^\dagger \overleftrightarrow D_\nu H\right)\left(H^\dagger \tilde W^{\mu\nu} H\right) \ ,
\ee
which indeed contain the terms in the right-hand side of \eqref{firstTermsNonLinear}. The terms above are of dimension 7 and are expected to be suppressed by the third power of the cutoff in the EFT (the third one is even of dimension 9, hence suppressed by the fifth power of the cutoff). One could then send (at least formally) the cutoff to infinity while keeping the decay constant $f$ fixed, thus decoupling those additional terms. However, we saw in section \ref{section:AbelianMatching} that, apart from the scale $f$ that always accompanies the axion, there is no other scale that suppresses the non-anomalous terms when they are obtained from heavy chiral matter, in other words chiral matter generates them as genuine dimension-5 terms. This is an example of non-decoupling effects when chiral matter is integrated out \cite{DHoker:1984izu,DHoker:1984mif,Feruglio:1994sv,Dudas:2009uq}, which make the use of a HEFT necessary. Consequently, the non-anomalous axion terms must be written using a non-linear representation of the SM gauge group. We comment further on this point in section \ref{axionCouplingsPheno}, when we discuss a precise example of a full realistic model. The bottom line is that the consistent choice is to use \eqref{termsToBeAdded} with a $\cO(1/f)$ suppression factor, which imples that the coefficient $c_i$ in \eqref{newTerms} are pure numbers, that do not disappear in any decoupling limit.
\section{Axions in chiral extensions of the SM}\label{eq:chiralextSM}

The formalism developed in the previous sections allows us to properly identify the axion couplings to massive gauge fields and has a relevant phenomenological application for the case of the SM-axion effective Lagrangian, when chiral fields (under the spontaneously broken gauge symmetries) are integrated out.

At tree level, in the UV theory, there are axion interactions with the SM particles. All of the latter are captured by the following SM-axion effective Lagrangian, keeping up to $d=5$ operators:
\begin{align}
\label{eq:effALPLag} 
\mathcal{L}^{d\leq 5}_{a} =& \, \frac{1}{2} (\partial_\mu a)^2 - \frac{1}{2} m_a^2 a^2 
+ \frac{\partial^\mu a}{f} \sum_{\psi} \bar \psi \, \mathbf{C}^{(\psi)} 
\gamma_\mu \psi \nonumber \\
&
- \frac{g_s^2 C_{GG}}{16\pi^2} \frac{a}{f} G \tilde G  
- \frac{g^2 C_{WW}}{16\pi^2} \frac{a}{f} W \tilde W 
- \frac{g'^2 C_{BB}}{16\pi^2} \frac{a}{f} B \tilde B \ ,   
\end{align}
where the sum over $\psi$ is taken over chiral fermions of the SM and $\mathbf C^{(\psi)}$ is a hermitian matrix in generation space. Here, the $C_{XX}$ coefficients are the anomaly coefficients of the PQ symmetry with the gauge group of gauge field $X$, when it is restricted to the SM fields only. 
Note that we work in the basis where the $d=5$ axion-Higgs operator 
$\partial^\mu a H^\dagger i \overleftrightarrow D_\mu H$ has been shifted away via an axion field redefinition \cite{Georgi:1986df}.

Since we are interested in axion couplings to gauge bosons, we could consider integrating out all SM fermions to end up with a bosonic EFT. However, in the SM, not even the top can safely be integrated out when considering the axion couplings to $W$/$Z$ bosons. Hence, if the axion couples to SM fields, one cannot rely on an EFT analysis when inspecting the axion couplings to massive gauge bosons and one should compute loop contributions in the full theory (see e.g. \cite{Bauer:2017ris}). On the other hand, an interesting EFT limit exists if we consider the possibility that axion couplings to $W$/$Z$ bosons are generated by extra 
fermions that are chiral under $SU(2)_L \times U(1)_Y$. SM chiral extensions do not admit a decoupling limit and are therefore severely constrained by Higgs couplings measurements, electroweak precision tests and direct searches. These measurements push the lower bound on the mass of these exotic states to several hundreds of GeV. Consequently, they can safely be integrated out when discussing the axion couplings to $W$/$Z$ bosons. 

In order to form the full SM-axion EFT below the mass scale of those new fermions, one should add to the couplings in \eqref{eq:effALPLag} the contributions of the heavy fermions which are integrated out. This brings additional terms, in particular additional couplings between the axion and the electroweak gauge fields which do not simply add up to the $C_{XX}$, as discussed in section \ref{sumrulesSection}. Henceforth, we focus on the contribution of the heavy chiral fields, which we identify later for a specific model, and we keep in mind when discussing phenomenology that the full amplitudes may include a loop contribution from the SM 
fermions in \eqref{eq:effALPLag}.
 
\subsection{Minimal SM chiral extensions}\label{section:minimalChiralBSM}
\label{sec:minimalchiral}

We wish to identify phenomenologically viable chiral extensions to SM that can provide a non-decoupling contribution to the SM-axion effective Lagrangian. To this end, we first stick to a single Higgs doublet\footnote{The extension to two Higgs doublets, relevant for the 
axion Lagrangian, will be discussed in section~\ref{sec:DFSZUV}.}
and we introduce a set of chiral fermions transforming irreducibly under $SU(3)_c\times SU(2)_L\times U(1)_Y$ as $(R_c, R_L, Y)$ and, following Ref.~\cite{Bizot:2015zaa}, we require the criteria:
\begin{enumerate}
\vspace{-5pt}
\item \label{crit:1} No massless fermion after EW symmetry breaking, apart for SM gauge singlets; 
\vspace{-7pt}
\item \label{crit:2} No gauge and Witten~\cite{Witten:1982fp} anomalies;
\vspace{-7pt}
\item \label{crit:3} Compatibility with Higgs coupling modifications; 
\vspace{-7pt}
\item \label{crit:4} No allowed bare mass terms.
\vspace{-5pt}
\end{enumerate}
If a bare mass term were to be allowed by the SM gauge symmetry, the mass of the fermions would be naturally heavier than the EW scale, thus enforcing an automatic decoupling in all observables, including the axion couplings to $W$/$Z$ bosons. So either the bare mass terms are accidentally suppressed compared to the EW scale or they have to be forbidden by extra discrete gauge symmetries. 

Condition \ref{crit:3}~restricts the choice of color representations to the trivial one, $R_c = 1$, so that SM Higgs production via gluon fusion 
is not affected by the new chiral sector. Focussing on renormalizable extensions, since the new chiral fermions need to become massive after EW symmetry breaking they need to couple to $H \sim (1,2,\tfrac{1}{2})$ via a Dirac-like Yukawa
\beq 
\bar f_L f_R H 
\quad \text{or} \quad 
\bar f_L f_R \tilde H 
\, ,  
\eeq
with $\tilde H = i \sigma_2 H^{*}$.
The quantum numbers of $f_L$ and $f_R$ can only be
\beq 
f_L \sim (1,2j_L + 1,Y) 
\, , \quad 
f_R \sim (1,2j_R + 1,Y-\tfrac{1}{2}) 
\quad \text{or} \quad 
f_R \sim (1,2j_R + 1,Y+\tfrac{1}{2}) 
\, , 
\eeq
with $\abs{j_L - j_R} = 1/2$. The minimal possibility is $j_L = 1/2$ and $j_R = 0$. Since the latter quantum numbers are reminiscent of the SM doublet and singlet leptons (for $Y=-\tfrac{1}{2}$), we denote them as\footnote{It is understood that doublets are LH and singlets are RH Lorentz spinors.}:
\beq 
L_1 \sim (1,2,Y) \, , \quad 
E_1 \sim (1,1,Y-\tfrac{1}{2}) 
\quad \text{or} \quad 
N_1 \sim (1,1,Y+\tfrac{1}{2}) 
\, . 
\eeq
Witten anomaly~\cite{Witten:1982fp} requires an even number of $SU(2)_L$ doublets, so we minimally introduce a second doublet 
\beq 
L_2 \sim (1,2,-Y) \, ,
\eeq 
where the hypercharge is fixed by the cancellation of the $SU(2)^2_L\times U(1)_Y$ anomaly.  At this point we need to consider $U(1)_Y$-gravitational and $U(1)^3_Y$ anomalies. The latter are already cancelled in the LH sector since $L_1 + L_2$ forms a vector-like pair whose bare mass has to be forbidden or suppressed to fulfill criteria~\ref{crit:4}. So we just need to pair the RH sector with a fermion with opposite-sign hypercharge: 
\begin{align} 
&L_1 \sim (1,2,+Y) \, , \quad 
E_1 \sim (1,1,+Y-\tfrac{1}{2}) 
\quad \text{or} \quad 
N_1 \sim (1,1,+Y+\tfrac{1}{2}) 
\, , \\
&L_2 \sim (1,2,-Y) \, , \quad 
E_2 \sim (1,1,-Y+\tfrac{1}{2}) 
\quad \text{or} \quad 
N_2 \sim (1,1,-Y-\tfrac{1}{2}) 
\, . 
\end{align}
Let us consider for instance the case with RH fermions $E_{1,2}$ (similar conclusions apply for the other choice). The Yukawas terms are
\beq
-\mathcal{L}_Y = y_{E_1} \bar L_1 E_1 H + y_{E_2} \bar L_2 E_2 \tilde H + \text{h.c.} \, . 
\eeq
After EW symmetry breaking only the $T^3_L = -\tfrac{1}{2} (+\tfrac{1}{2})$ component of the $L_{1}$ ($L_{2}$) doublet with electric charge $Q_{1} = -\tfrac{1}{2} + Y$ ($Q_{2} = +\tfrac{1}{2} - Y$) picks up a mass, while the $T^3_L = +\tfrac{1}{2}(-\tfrac{1}{2})$ component with  $Q_{1} = +\tfrac{1}{2} + Y$ ($Q_{2} = -\tfrac{1}{2} - Y$) is unpaired and hence, remains massless. For $Y = -1/2$ the massless chiral fermions are electrically neutral, but they still retain an interaction with the $Z$ boson and hence are ruled out. In conclusions, we need to include both $E_{1,2}$ and $N_{1,2}$ pairs in order to not have massless fermions. Therefore the minimal setup which fulfills criteria ~\ref{crit:1}--\ref{crit:4} is\footnote{The case $Y=0$ needs to be discussed separately, 
but it does not lead to a more minimal setup. 
In principle, one could consider the 
anomaly free content 
$L_1 \sim (1,2,0)$,  
$E_1 \sim (1,1,-\tfrac{1}{2})$, 
$N_1 \sim (1,1,+\tfrac{1}{2})$, 
$L_2 \sim (1,2,0)$.
However, in such case the Yukawa sector is 
$-\mathcal{L}_Y 
= y^\alpha_{E_1} \bar L_\alpha E_1 H 
+ y^\alpha_{N_1} \bar L_\alpha N_1 \tilde H + \text{h.c.}$,  
where $\alpha=1,2$ is a flavour index. 
Without loss of generality with can do a $U(2)$ flavour transformation 
in such a way to align the $U(2)$ vector such that $y_{E_1} \propto (0,1)$. 
Hence the $Q = T^3_L =1/2$ component of $L_1$ remains massless, 
in contradiction with point \ref{crit:1}.} 
\begin{align} 
\label{eq:chiralfam1}
&L_1 \sim (1,2,+Y) \, , \quad 
E_1 \sim (1,1,+Y-\tfrac{1}{2}) 
\, , \quad 
N_1 \sim (1,1,+Y+\tfrac{1}{2}) 
\, , \\
\label{eq:chiralfam2}
&L_2 \sim (1,2,-Y) \, , \quad 
E_2 \sim (1,1,-Y+\tfrac{1}{2}) 
\, , \quad  
N_2 \sim (1,1,-Y-\tfrac{1}{2}) 
\, . 
\end{align}
The Yukawa sector is then 
\beq
\label{eq:Yuk1H}
-\mathcal{L}_Y = 
y_{E_1} \bar L_1 E_1 H + y_{E_2} \bar L_2 E_2 \tilde H +
y_{N_1} \bar L_1 N_1 \tilde H + y_{N_2} \bar L_2 N_2 H + \text{h.c.} \, , 
\eeq
and all chiral fermions pick up a mass after EW symmetry breaking.

Note that gauge anomaly cancellation is immediate since the field content in \eqref{eq:chiralfam1}-\eqref{eq:chiralfam2} is vector-like with respect to the SM gauge group. Nonetheless, as we said previously, we still consider this field content as chiral since we do not write bare mass terms and only use EWSB to produce the heavy masses. This leads to a theory where the massive states couple asymetrically when projected onto their LH or RH components, which is what we took as a definition for ``chiral''. We can remain agnostic about the reason that makes the bare masses suppressed with respect to the couplings to the Higgs field. However, note that bare mass terms can be forbidden by using discrete gauge symmetries, for instance a $\mathbb{Z}_2$ symmetry under which $L_1,E_1,N_1$ and all SM leptons are odd, and the other fields even, is anomaly-free and sufficient to forbid bare mass terms, while allowing for all the Yukawa couplings. Considering such a discrete gauge symmetry makes the spectrum genuinely chiral.

\subsection{A DFSZ-like UV completion}
\label{sec:DFSZUV}

In order to host the axion field as the Goldstone of a spontaneously broken 
$U(1)_\text{PQ}$ symmetry, we need to properly extend the scalar sector. 
We here provide a renormalizable UV completion of the SM-axion 
Lagrangian which is inspired by the standard DFSZ \cite{Zhitnitsky:1980tq,Dine:1981rt} 
axion. The main difference is that we do not require the axion to solve the strong CP 
problem; we improperly keep the label PQ, even though the axion does not necessarily has 
a QCD anomaly,
and we do not forbid that there is an extra source of $U(1)_\text{PQ}$ breaking giving an 
axion mass that is unrelated to the axion decay constant. However, in what follows we do not discuss any explicit source of PQ breaking, and we derive the axion couplings from a lagrangian that has an exact PQ symmetry. Indeed, the axion mass is irrelevant for our one-loop computations with fermion lines.

The scalar sector comprises a complex SM singlet 
$\Phi$ and two Higgs doublets $H_{1,2} \sim (1,2,+\frac{1}{2})$, 
with scalar potential 
\beq
\label{eq:VDFSZ}
V(H_1, H_2, \Phi) = 
V_{\rm r.i.} (\abs{H_1}, \abs{H_2}, \abs{\Phi}, |H_1^\dag H_2|) + 
\lambda \, H_1^\dag H_2 \Phi^2 + \text{h.c.} \, , 
\eeq
which contains all the 
re-phasing invariant terms allowed by gauge invariance 
plus a non-hermitian 
operator which is responsible for the explicit breaking 
of the three re-phasing symmetries $U(1)_{\Phi} \times U(1)_{H_1} \times U(1)_{H_2}$ 
into two linearly independent $U(1)$'s, to be identified with 
$U(1)_Y\times U(1)_{\rm PQ}$. \eq{eq:VDFSZ} implies 
\beq 
X_1 - X_2 = 2 \, ,
\eeq
where $X_{1,2}$ denote the PQ charges of $H_{1,2}$, 
and we have normalized $X_{\Phi} = 1$. 

It turns out that with a single Higgs doublet in the Yukawa Lagrangian of the exotic fermions
(cf.~\eq{eq:Yuk1H}), there are no dimension-5 axion-gauge bosons couplings in the EFT below the exotics mass, as we show in appendix \ref{appendixPheno}. In order to obtain an EFT with non-trivial axion couplings, 
we modify \eq{eq:Yuk1H} in the following way
\beq
\label{eq:Yuk2H}
- \mathcal{L}_Y = 
y_{E_1} \bar L_1 E_1 H_1 + y_{E_2} \bar L_2 E_2 \tilde H_2 +
y_{N_1} \bar L_1 N_1 \tilde H_2 + y_{N_2} \bar L_2 N_2 H_1 + \text{h.c.} \, , 
\eeq
which implies the following constraints on the $U(1)_{\rm PQ}$ charges: 
\be
\bead 
& -X_{L_1} + X_{E_1} + X_1 = 0 \, , \qquad  -X_{L_1} + X_{N_1} - X_2 = 0 \, , \\
& -X_{L_2} + X_{E_2} - X_2 = 0 \, , \qquad  -X_{L_2} + X_{N_2} + X_1 = 0 \, .
\eead
\label{PQrelations}
\ee
We recall that, although gauge symmetry allows the bare mass terms $L_1 L_2$, $E_1 E_2$ and $N_1 N_2$, they can be forbidden, e.g.~via a discrete gauge symmetry, in order to avoid decoupling effects. Note also that we could have chosen different assignments of $H_{1,2}$ in \eqref{eq:Yuk2H}, which amount to different choices of PQ symmetry, and induce different axion and/or Higgs boson phenomenology. Henceforth, we stick to this choice and briefly discuss other cases in appendix \ref{appendixPheno}. We do not specify the Higgs fields assignment in the SM Yukawa couplings, we simply assume that it only involves $H_{1,2}$ and no third Higgs doublet, and that it is done such that there are no tree-level flavour changing neutral currents. One possibility is to couple them as in the original DFSZ model, in which case \eqref{eq:effALPLag}  is simply the usual DFSZ axion-SM EFT. As we said, we focus here on the contributions from the extra heavy fermions.

Next, we proceed to identify the axion Goldstone mode in terms of the scalar components
\beq
\label{eq:axiondirDFSZ}
H_1 \supset 
\frac{v_1}{\sqrt{2}} e^{i\frac{a_1}{v_1}} 
\, \binom{0}{1}
\ , \quad
H_2 \supset 
\frac{v_2}{\sqrt{2}} e^{i\frac{a_2}{v_2}} 
\,
\binom{0}{1}
\ , \quad
\Phi \supset 
\frac{v_\Phi}{\sqrt{2}} e^{i\frac{a_\Phi}{V_\Phi}}
\ , 
\eeq
where we have neglected EM-charged and radial modes that have 
no projection on the axion field denoted as $a$. 
In order to identify the latter in terms of  $a_{1,2,\Phi}$  let us write down the 
classically conserved PQ current, restricted to the scalar sector 
\begin{align} 
\label{eq:JPQDFSZ}
J^{\rm PQ}_{\mu} &= 
-\Phi^\dag i \overset\leftrightarrow{\partial_\mu} \Phi 
-X_{1} H_1^\dag i \overset\leftrightarrow{\partial_\mu} H_1
-X_{2} H_2^\dag i \overset\leftrightarrow{\partial_\mu} H_2 
+ \ldots \nonumber \\
&= V_{\Phi} \partial_\mu a_\Phi + X_1 v_1 \partial_\mu a_1 + X_2 v_2 \partial_\mu a_2 + \ldots \,, 
\end{align}
where we only included the scalar terms relevant for the identification of the axion. 
Following the Goldstone theorem $\langle 0 | J^{\rm PQ}_{\mu}  | a \rangle \sim i f p_\mu$, 
the axion-Goldstone field is defined as
\beq 
\label{eq:defaxionDFSZ}
a = \frac{1}{f} 
\( V_{\Phi} a_\Phi + X_1 v_1 a_1 + X_2 v_2 a_2 \) \,, 
\quad 
f^2 = V_\Phi^2 + X_1^2 v^2_1 + X_2^2 v^2_2 \,, 
\eeq
so that $J^{\rm PQ}_{\mu} \supset f \partial_\mu a$. 
Under a PQ transformation 
$a_{1,2} \to a_{1,2} + \kappa X_{1,2} v_{1,2}$ and 
$a_\Phi \to a_\Phi + \kappa v_\Phi$,   
the axion field transforms as $a \to a + \kappa f$. 

Requiring that under a $U(1)_Y$ gauge transformation 
the axion field $a$ remains invariant 
yields $X_1 Y_1 v_1^2 + X_2 Y_2 v_2^2 = 0$, where $Y_{1,2}= 1/2$ are the hypercharges of the Higgs doublets $H_{1,2}$.  
Hence, all the PQ charges in the scalar sector are fixed in terms of $\tan\beta = v_2 / v_1$: 
\beq 
\label{eq:DFSZHiggsX}
X_{\Phi} = 1 \, , \qquad 
X_1 = 2 \sin^2\beta \, , \qquad 
X_2 = -2 \cos^2\beta \, , 
\eeq
where we have defined $v_1/v = \cos\beta$, $v_2/v = \sin\beta$, 
with $v = \sqrt{v_1^2 + v_2^2} \simeq 246$ GeV. Substituting these expressions into \eq{eq:defaxionDFSZ} we obtain: 
\beq
\label{eq:acanonDFSZ}
f^2 = V_\Phi^2 + v^2 (\sin2\beta)^2\, . 
\eeq
In the limit $V_\Phi \gg v$ we can approximate $f \simeq V_\Phi$. 

The axion coupling to the new chiral fermions 
can be derived by 
inverting the first relation in \eq{eq:defaxionDFSZ}
to express  $a_{1,2}$ in terms of $a$. This boils down to replace  
$a_1 / v_1 \to X_{1} a / f$, 
$a_2 / v_2 \to X_{2} a / f$ in \eq{eq:axiondirDFSZ} 
and yields, decomposing the doublets as $L_{1} = (N_{L_{1}},E_{L_{1}})^T$, $L_{2} = (E_{L_{2}},N_{L_{2}})^T$,
\begin{align}
\label{linearBasisUVmodel}
- \mathcal{L}_Y &\supset 
m_{E_1} \(e^{i  X_1 \frac{a}{f}} \)  \bar E_{L_1} E_1 
+ m_{E_2}  \(e^{-i  X_2 \frac{a}{f}} \)  \bar E_{L_2} E_2 \nonumber \\
&+ m_{N_1} \(e^{-i  X_2 \frac{a}{f}} \)  \bar N_{L_1} N_1 
+ m_{N_2} \(e^{i  X_1 \frac{a}{f}} \)  \bar N_{L_2} N_2 
+ \text{h.c.} \, , 
\end{align}
where we defined the Dirac mass terms
$m_{E_{1,2}} = y_{E_{1,2}} \tfrac{v_1}{\sqrt{2}}$ and 
$m_{N_{1,2}} = y_{N_{1,2}} \tfrac{v_2}{\sqrt{2}}$.
We could go further and remove the axion field from the mass terms  
by redefining the fermion fields via a field-dependent chiral transformation. The non-invariance of the fermion kinetic terms plus possible anomalous transformations 
lead in turn to an axion effective Lagrangian similar to \eqref{eq:effALPLag}. However, it is easier to obtain the axion-gauge bosons couplings in the basis of \eqref{linearBasisUVmodel}. The EFT terms that arise due to anomalous transformations can be found in appendix \ref{appendixPheno}.

\subsection{Axion couplings to gauge bosons and sum-rules}\label{axionCouplingsPheno}

We now derive the axion EFT below the mass of the new fermions. For that, we define the massive eigenstates $N_i=N_i+N_{L_i}$, $E_i=E_i+E_{L_i}$ and extract their gauge couplings from
\begin{eqnarray}
\cL\supset&&\overline{L_i}\gamma^\mu\left(g\frac{\sigma^a}{2}W^a_\mu+g'(-1)^{i+1}YB_\mu\right)L_i+g'\overline{N_i}\gamma^\mu (-1)^{i+1}\left(Y+\frac{1}{2}\right)B_\mu N_i\nonumber \\
&&+g'\overline{E_i}\gamma^\mu (-1)^{i+1}\left(Y-\frac{1}{2}\right)B_\mu E_i \ .
\end{eqnarray}
Assuming equal masses within a $SU(2)$ doublet for simplicity ($m_{N_i}=m_{E_i}$) and using the formulae in appendix \ref{appendixPheno}, the axion couplings in the EFT read
\be
\cL\supset-g'^2\frac{(1+12Y^2)(X_1-X_2)}{96\pi^2}\frac{a}{f}B\tilde B - g^2\frac{X_1-X_2}{96\pi^2}\frac{a}{f}W^a\tilde W^a\big\vert_\text{lin.}- gg'\frac{X_1-X_2}{96\pi^2}\frac{a}{f}B\tilde W^3\big\vert_\text{lin.} \ ,
\label{actualAxionVectorTerms}
\ee
where we used the same letter to refer to the gauge fields and to their field strengths, and where $W^a_{\mu\nu}\vert_\text{lin}=\partial_\mu W^a_\nu-\partial_\nu W^a_\mu$ (we only compute couplings of an axion to two gauge fields, which are only sensitive to the linear parts of the field strengths). Although we normalized $X_1-X_2=2$ previously, we kept it in \eqref{actualAxionVectorTerms} to make the charge dependence explicit.

Let us notice that the last term cannot be reproduced by a UV PQ anomaly, since there is no non-vanishing $U(1)_\text{PQ}\times U(1)_Y \times SU(2)_W$ anomaly coefficient. In particular, it does not match the usual ansatz \eqref{SManomalousOperators}, but it can be obtained including also a combination of the EFT terms in \eqref{termsToBeAdded}\footnote{It can be checked that one finds $c_3=0$ when matching between \eqref{actualAxionVectorTerms} onto \eqref{SManomalousOperators} and \eqref{termsToBeAdded}. This is due to the fact that $c_3$ violates custodial symmetry, while we used the simplifying custodial symmetry limit ($m_{N_i}=m_{E_i}$) in our computation, which is also motivated by electroweak precision tests, as discussed later.}. Therefore, the present UV example is a confirmation of a statement we made previously: the terms in \eqref{termsToBeAdded} appear at dimension 5 in the lagrangian, and are not suppressed by the mass of the heavy fermions.

We can express \eqref{actualAxionVectorTerms} in terms of vector mass eigenstates, focusing on neutral bosons,
\be
\bead
-\frac{16\pi^2}{e^2}\cL\ \supset \ &(X_1-X_2)\left(2Y^2+\frac{1}{2}\right)\frac{a}{f}F\tilde F+(X_1-X_2)\left(-t_W\left(4Y^2+\frac{1}{2}\right)+\frac{t_W^{-1}}{2}\right)\frac{a}{f}F\tilde Z\\
&+(X_1-X_2)\left(t_W^2\left(2Y^2+\frac{1}{6}\right)-\frac{1}{6}+\frac{t_W^{-2}}{6}\right)\frac{a}{f}Z\tilde Z \ ,
\label{axionPhotonZcouplings}
\eead
\ee
It is evident that these couplings do not derive solely from UV anomalies by noticing that the sum rules in \eqref{sumRuleEFTCoeffs} are violated:
\be
C_{\gamma\gamma}+s_W^{-2}(1-t_W^2)C_{Z\gamma}-\frac{1}{s_W^2c_W^2} C_{ZZ}=\frac{X_1-X_2}{12c_W^2s_W^2}\ ,\quad C_{\gamma\gamma}+s_W^{-2}C_{Z\gamma}-(1+t_W^{-2})C_{WW}=\frac{X_2-X_1}{12s_W^2} \ .
\ee
That means that one cannot define any PQ symmetry whose UV anomalies reproduce \eqref{axionPhotonZcouplings} (see appendix \ref{appendixPheno} for more details\footnote{As we discuss in appendix \ref{appendixPheno}, it is possible to understand the precise coefficients in \eqref{axionPhotonZcouplings} by treating the $Z$ boson as a massive abelian gauge boson, following the prescription at the end of section \ref{section:AbelianMatching}.}). This breakdown of the sum rules, that can be directly tested given an observation of the axion-gauge boson couplings, is a smoking gun of the presence of a chiral heavy sector charged under the PQ symmetry. 

\begin{figure}[t!]
\centering
\includegraphics[width=0.75\textwidth]{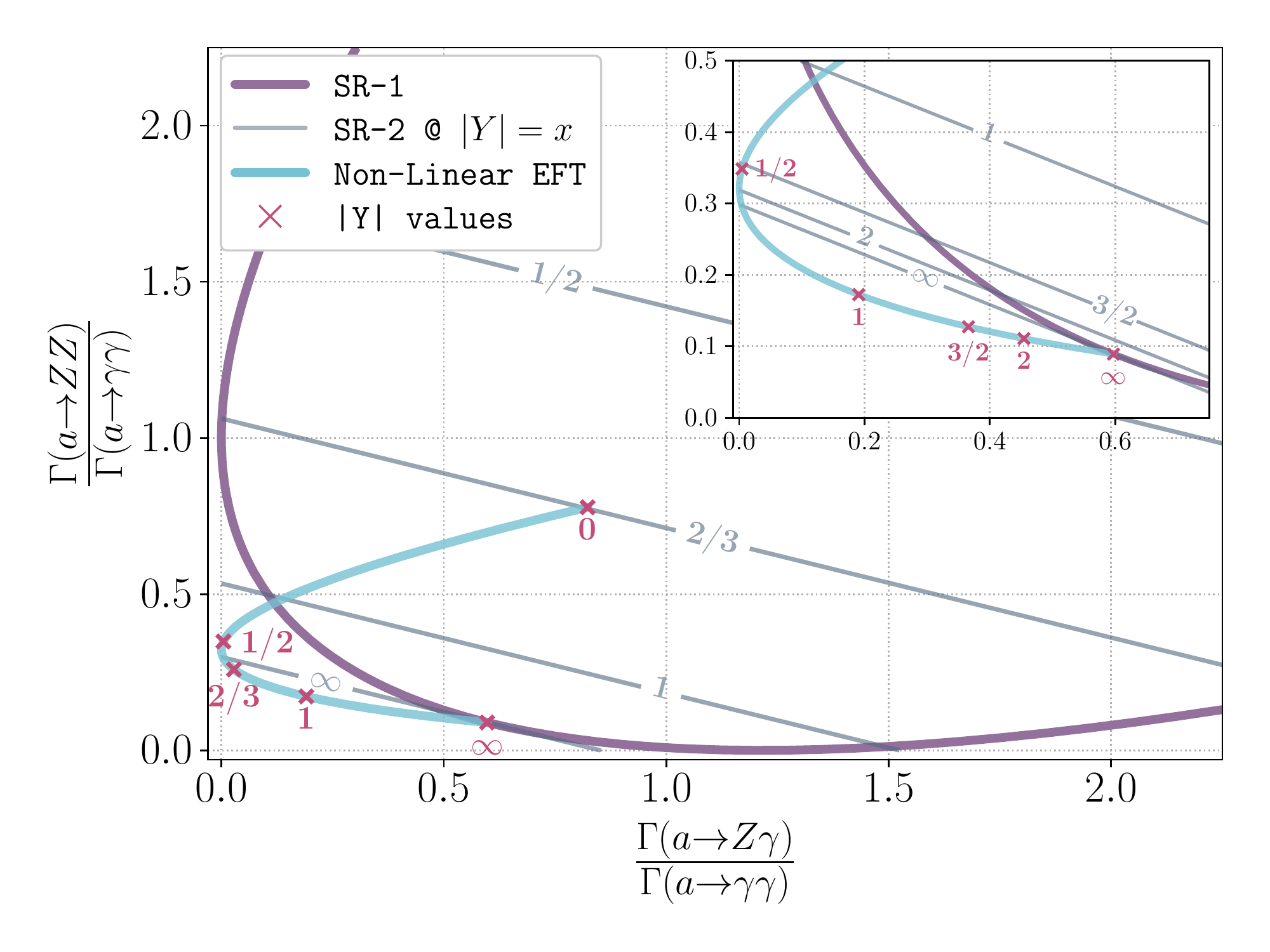}
\caption{\it 
Plot of the ratio of decay rates $\frac{\Gamma(a\rightarrow ZZ)}{\Gamma(a\rightarrow \gamma\gamma)}$ as a function of the ratio of decay rates $\frac{\Gamma(a\rightarrow Z\gamma)}{\Gamma(a\rightarrow \gamma\gamma)}$, assuming that the decays are kinematically allowed. The anomaly-based sum rule SR-1, given in \eqref{eq:sumrule}, corresponds to the purple curve. The sum rule SR-2, given in \eqref{eq:sumruleBis}, can only be traced on a plane if a ratio of partial rates is fixed. We therefore fix $\frac{\Gamma(a \to WW)}{\Gamma(a\to\gamma\gamma)}$ to its value derived from \eqref{axionPhotonZcouplings} as a function of $Y$. Each gray line then corresponds to the linear relation in \eqref{eq:sumruleBis} when we scan values of $Y$. Finally, the blue curve uses the explicit couplings in \eqref{axionPhotonZcouplings}.}
\label{fig:sumRuleViolationModel}
\end{figure}

It is useful to reformulate the sum rules in \eqref{sumRuleEFTCoeffs} in terms of observable quantities, for instance in terms of partial decay rates of the axion. Assuming that $m_a > 2m_Z$ so that all decays are allowed, and ignoring the SM contributions, the rates read (see e.g.~\cite{Franceschini:2016gxv})
\be
\bead
\Gamma(a \to \gamma\gamma) = C^2_{\gamma\gamma} \frac{\alpha^2 m^3_a}{64 \pi^3 f^2} \, &, \quad \Gamma(a \to WW) = C^2_{WW} \frac{\alpha^2 m^3_a}{32 \pi^3 s^4_W f^2} \, , \\
\Gamma(a \to ZZ) = C^2_{ZZ} \frac{\alpha^2 m^3_a}{64 \pi^3 s^4_W c^4_W f^2} \, &, \quad \Gamma(a \to Z\gamma) = C^2_{Z\gamma} \frac{\alpha^2 m^3_a}{32 \pi^3 s^2_W c^2_W f^2} \, ,
\eead
\ee
where the couplings $C_{XX}$ are defined in \eqref{SManomalousOperatorsZphotonW}. Using the first identity in \eqref{sumRuleEFTCoeffs}, one sees that the following holds
\be
\label{eq:sumrule}
\text{SR-1: }\ \left[\frac{\Gamma(a \to ZZ)}{\Gamma(a\to\gamma\gamma) } - 1
- \frac{\(t_W^2 -1 \)^2}{2t_W^2} \frac{\Gamma(a \to Z\gamma)}{\Gamma(a\to\gamma\gamma) }\right]^2 -\frac{2\(t_W^2 -1 \)^2}{t_W^2}\frac{\Gamma(a \to Z\gamma)}{\Gamma(a\to\gamma\gamma)}= 0 \, ,
\ee
which is a relation between two quantities that can be traced on a plane. Another sum rule that follows from both identities in \eqref{sumRuleEFTCoeffs} is
\be
\text{SR-2: }\ \Gamma(a\to\gamma\gamma) 
+ \frac{1}{2} \(t^{-2}_W - 1\) \Gamma(a \to WW) 
- t^{-2}_W  \Gamma(a \to ZZ) 
+ \frac{1}{2} \(1 - t^{-2}_W \) \Gamma(a \to Z\gamma) = 0 \, .
\label{eq:sumruleBis}
\ee
By performing a low-energy measurement, one can test those 
sum-rules; if at least one of them is violated, we can 
conclude that the fermionic UV completion is chiral, whereas a vector-like one (e.g.~a KSVZ-like model) always satisfies them. The violation of the sum rules in our specific model is displayed in figure \ref{fig:sumRuleViolationModel}, where one can check that the model only satisfies both sum rules when $\abs{Y}=\infty$.

We should stress the caveat that SM loop contributions also violate the naive sum rules, so they should be taken into account. In particular, the axion-fermion couplings in \eqref{eq:effALPLag}, which are responsible for the SM loop corrections to the sum rule, should be reconstructed from low-energy data in order to extract the bosonic EFT terms on which the sum rule can be tested. 

\subsection{Phenomenology of SM chiral extensions} 

Due to their non-decoupling nature, SM chiral extensions have 
an interesting phenomenology that we summarize in the following. 
The aim of this Section is to show that the minimal chiral setup 
in \sect{sec:minimalchiral} is strongly constrained, but not yet ruled-out.

\subsubsection{Electroweak precision tests} 

The contribution of the new exotic fermions in \eqs{eq:chiralfam1}{eq:chiralfam2}  
to the $S$ and $T$ parameters
is \cite{Bizot:2015zaa}
\begin{align}
S&= \frac{1}{6\pi} 
\[ \(1 - 2 Y \log\frac{m^2_{N_1}}{m^2_{E_1}}\) 
+ \(1 + 2 Y \log\frac{m^2_{N_2}}{m^2_{E_2}}\) 
+ \mathcal{O}\(\frac{m^2_{Z}}{m^2_{N,\, E}}\)
\] 
\approx \frac{1}{3\pi} \, , \\
T&= \frac{1}{16\pi c^2_W s^2_W m^2_Z} 
\( m^2_{N_1} + m^2_{E_1} 
- 2 \frac{m^2_{N_1}m^2_{E_1}}{m^2_{N_1}-m^2_{E_1}} \log\frac{m^2_{N_1}}{m^2_{E_1}}\) \nonumber \\
&+ \frac{1}{16\pi c^2_W s^2_W m^2_Z} 
\( m^2_{N_2} + m^2_{E_2} 
- 2 \frac{m^2_{N_2}m^2_{E_2}}{m^2_{N_2}-m^2_{E_2}} \log\frac{m^2_{N_2}}{m^2_{E_2}}\) 
\approx 0
\, ,
\end{align}
where the approximation in the last steps 
holds in the custodial limits 
$y_{N_1}=y_{E_1}$ and $y_{N_2}=y_{E_2}$. 
Recent fits for oblique parameters, e.g.~from Gfitter \cite{Gfitter}, yield
\beq 
S= 0.05 \pm 0.11 \, , \qquad 
T= 0.09 \pm 0.13 \, ,  
\eeq
which are easily satisfied in the custodial limit.

\subsubsection{Higgs couplings}\label{HiggsCouplingsSection}

We now study the constraints from Higgs couplings measurements. In particular, we assess the impact of the new heavy fermions on the decay rate of the Higgs boson to two photons, or to a photon and a Z boson.

The useful formulae are the following. Consider a fermion $\psi$ of mass $m_\psi$ with couplings given by the following Lagrangian,
 \be
\cL_\psi=\overline{\psi}(i\slashed{\partial}- m_\psi)\psi-\frac{x_\psi m_\psi}{v}h\overline{\psi}\psi+eQ_\psi\overline{\psi}\gamma^\mu\psi A_\mu+\frac{e}{c_Ws_W}\overline{\psi}\gamma^\mu\left(\frac{T^3_\psi}{2}-Q_\psi s_W^2-\frac{T^3_\psi}{2}\gamma_5\right)\psi Z_\mu \ ,
\label{lagHiggsCouplings}
\ee
where $h$ is the $125$ GeV Higgs, $A_\mu$ and $Z_\mu$ the photon and $Z$ boson fields, $T^3_\psi$ is the eigenvalue of the third generator of $SU(2)_L$ when it acts on the left-handed component of $\psi$ ($\psi_L=\frac{1-\gamma^5}{2}\psi$), so that $T^3_\psi=\pm\frac{1}{2}$ when $\psi_L$ arises from a doublet in the fundamental of $SU(2)_L$, and $x_\psi$ is a number. Its one-loop contributions to the amplitudes $h\to\gamma\gamma$ and $h\to\gamma Z$ are \cite{Djouadi:2005gi}
\be
\cA^\psi_{\gamma\gamma}\approx\frac{4}{3}x_\psi Q_\psi^2 \ , \quad \cA^\psi_{Z\gamma}\approx-\frac{1}{3}x_\psi Q_\psi\frac{T^3_\psi-2Q_\psi s_W^2}{c_W} \ ,
\label{higgsAmplitudeFormulae}
\ee
where we assumed that $\psi$ is much heavier than the Higgs and the $Z$ boson, which holds for the heavy fermions we consider here. In the SM, these amplitudes are dominated by the loop of the $W$ gauge boson interfering negatively with the loop of the top quark and they amount to $A_{\gamma\gamma}^{\rm SM} \approx -6.5$ and $\mathcal{A}_{\gamma Z}^{\rm SM} 
\approx 5.7$ at LO (the dominant QCD NLO corrections give a correction of order 5\%).

We can apply these formulae to the case of our model in \eqref{eq:Yuk2H}. However, the quantity $x_\psi$ in \eqref{lagHiggsCouplings} depends on the entries of the scalar potential of the two Higgs doublets. Indeed, the generic vacuum of a 2HDM is parametrized by two angles $\alpha,\beta$ (see for instance \cite{Branco:2011iw} for a review). For a $SU(2)_L$ doublet $Q_L=(u_L,d_L)$ and a singlet $d_R$, the Yukawa couplings to the light Higgs, which we choose to be the known $125$ GeV particle for definiteness, read,
\be
\overline{Q_L}d_RH_1+h.c.\supset m_d\left(1-\frac{\sin\alpha}{\cos\beta}\frac{h}{v}\right)\overline{d}d \ , \quad \overline{Q_L}d_R H_2+h.c.\supset m_d\left(1+\frac{\cos\alpha}{\sin\beta}\frac{h}{v}\right)\overline{d}d \ ,
\ee
and the couplings of $u_L$ to a singlet $u_R$ are similarly obtained. Couplings of the light Higgs to the vector bosons are given by their SM values times $\sin(\beta-\alpha)$.

\begin{figure}[ht!]
\centering
\includegraphics[width=0.75\textwidth]{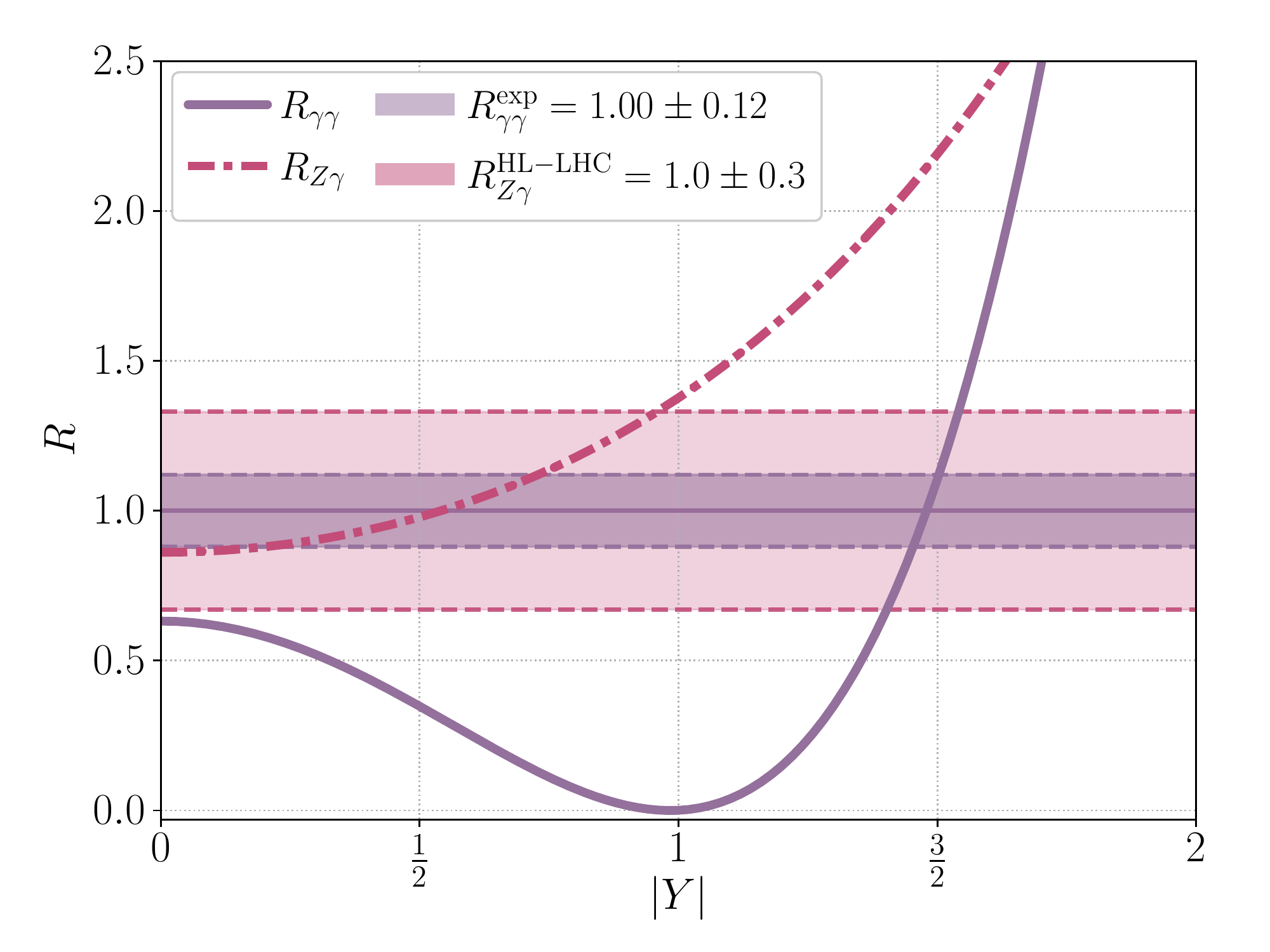}
\caption{\it
$R_{\gamma\gamma}$ and $R_{Z\gamma}$ as a function of $\abs{Y}$ for the model in \eqref{eq:Yuk2H} in the alignement limit. The horizontal solid and dashed purple lines show the experimental value of $R_{\gamma\gamma}$ and the 1$\sigma$ boundaries, respectively. The pink ones correspond to projections of $R_{Z\gamma}$ at HL-LHC.}
\label{fig:RggvsY}
\end{figure}

To get a SM-like light Higgs, one can go to the alignment limit $\beta-\alpha=\pi/2$, in which case the couplings to vector bosons is SM-like, as well as the Yukawa couplings since then $-\sin\alpha/\cos\beta=\cos\alpha/\sin\beta=1$. The Higgs signals are therefore modified as if there was a single Higgs doublet, namely as if the model was given by the Yukawa couplings in \eq{eq:Yuk1H}. One must then use $x_\psi=1$ in \eqref{lagHiggsCouplings} for all the heavy fermions, and this yields $\mathcal{A_{\gamma\gamma}^{\rm new}} 
\approx \frac{4}{3} (1 + 4Y^2)$. Writing the modified Higgs width to photons as 
\beq 
R_{\gamma\gamma} = \frac{\abs{\mathcal{A_{\gamma\gamma}^{\rm SM}} + \mathcal{A_{\gamma\gamma}^{\rm new}}}^2}{\mathcal{\abs{A_{\gamma\gamma}^{\rm SM}}}^2} \, , 
\eeq
a recent ATLAS analysis finds $R_{\gamma\gamma} = 1.00 \pm 0.12 $~\cite{Aad:2019mbh}. 
There is only the possibility that the new contribution interferes negatively with the SM 
amplitude, namely $\mathcal{A_{\gamma\gamma}^{\rm new}} \approx -2 
\mathcal{A}_{\gamma\gamma}^{\rm SM} \approx 13.0$. In such a case the allowed 
2$\sigma$ range is $1.43 \lesssim \abs{Y} \lesssim 1.53$ (cf.~\fig{fig:RggvsY}). 

A correlated signal in the $\gamma Z$ channel is $\mathcal{A}_{\gamma Z}^{\rm new} 
\approx -\frac{2}{3}c_W [1 - (1 + 8Y^2) t^2_{W}]$, leading to a large deviation in the region where $\abs{Y} \approx 1.5$ compatible with the diphoton channel (cf.~\fig{fig:RggvsY}). The $\gamma Z$ decay channel of the Higgs has not been observed yet and HL-LHC is expected to measure $\kappa_{\gamma Z}$ within a $10\%$ precision~\cite{Cepeda:2019klc}. Future lepton colliders would not offer any improvement on that bound and only FCC-hh could reach a precision below $1\%$ \cite{deBlas:2019rxi}.

The alignment limit of the model in \eqref{eq:Yuk2H} therefore predicts a strong departure of $R_{Z\gamma}$ from its SM-value. However, this conclusion can be evaded in other limits of the 2HDM. In particular, the modifications to the Higgs signals can be strongly suppressed in the so-called wrong-sign limit \cite{Ginzburg:2001wj,Gunion:2002zf,Ferreira:2014naa}. This limit uses the fact that only the sign of the top Yukawa has been measured. That means that, defining the labelling such that the top quark couples to $\tilde H_2$, we must enforce $\cos\alpha=\sin\beta\implies\alpha=\pm\left(\frac{\pi}{2} -\beta\right)$. The minus sign gives the alignment limit, but there exists another viable option: the plus sign gives the wrong-sign limit, $\alpha+\beta=\frac{\pi}{2}$, so called because in this limit, the bottom-Higgs coupling is $-m_b/v$, i.e. $-1$ times its SM value. The gauge bosons couplings of the Higgs are not SM-like either, and are 
\be
\sin(\beta-\alpha)=\sin(2\beta-\frac{\pi}{2})=-\cos(2\beta)=-\frac{\cos^2\beta-\sin^2\beta}{\cos^2\beta+\sin^2\beta}=\frac{\tan^2\beta-1}{\tan^2\beta+1}
\ee
times their SM values. At large $\tan\beta$, this goes to $1$, but differs from it at fixed $\tan\beta$. Current measurements of the Higgs couplings then impose $\tan\beta\gtrsim 4$ at 68\% C.L.~\cite{ATLAS:2020qdt}, while the limit is expected to increase to $\tan\beta\gtrsim 12$ at HL-LHC~\cite{Cepeda:2019klc}. In this limit, heavy fermions coupled to $H_1$ have $x_\psi=-1$ in \eqref{lagHiggsCouplings}, whereas those coupled to $H_2$ have $x_\psi=1$, so that cancellations are possible in \eqref{higgsAmplitudeFormulae}. For instance, in the case of our model in \eqref{eq:Yuk2H} the partial amplitudes in \eqref{higgsAmplitudeFormulae} simply vanish, and the modifications to the Higgs signals are those of a 2HDM in the wrong-sign limit without extra chiral matter.

\subsubsection{Stable charged particles and direct searches} 

Except for $\abs{Y} = 3/2$, the exotic leptons do not mix the SM ones and the lightest state of the spectrum 
is electrically charged 
($Q = Y \pm 1/2$) and stable due to 
exotic lepton number (which arises accidentally due to $Y$). 
Charged relics are cosmologically dangerous and largely excluded. To avoid 
cosmological problems one has to invoke low-scale inflation 
$\text{max} \{H_I,T_{\rm RH}\} \lesssim$ TeV, such that charged relics are 
either diluted by inflation or never thermally produced. 
On the other hand, stable charged particles yield striking signatures at colliders 
in the forms of charged track, anomalous energy loss in calorimeters, longer time of flights, etc. 
Current bounds for $Q \lesssim 2$ are of order $700$ GeV \cite{Khachatryan:2016sfv}, 
which correspond to an electroweak contribution to their mass with Yukawas which saturate the perturbative unitarity limit. 

\begin{figure}
	\centering
	\includegraphics[trim= 10 10 20 10, clip, width=0.49\linewidth]{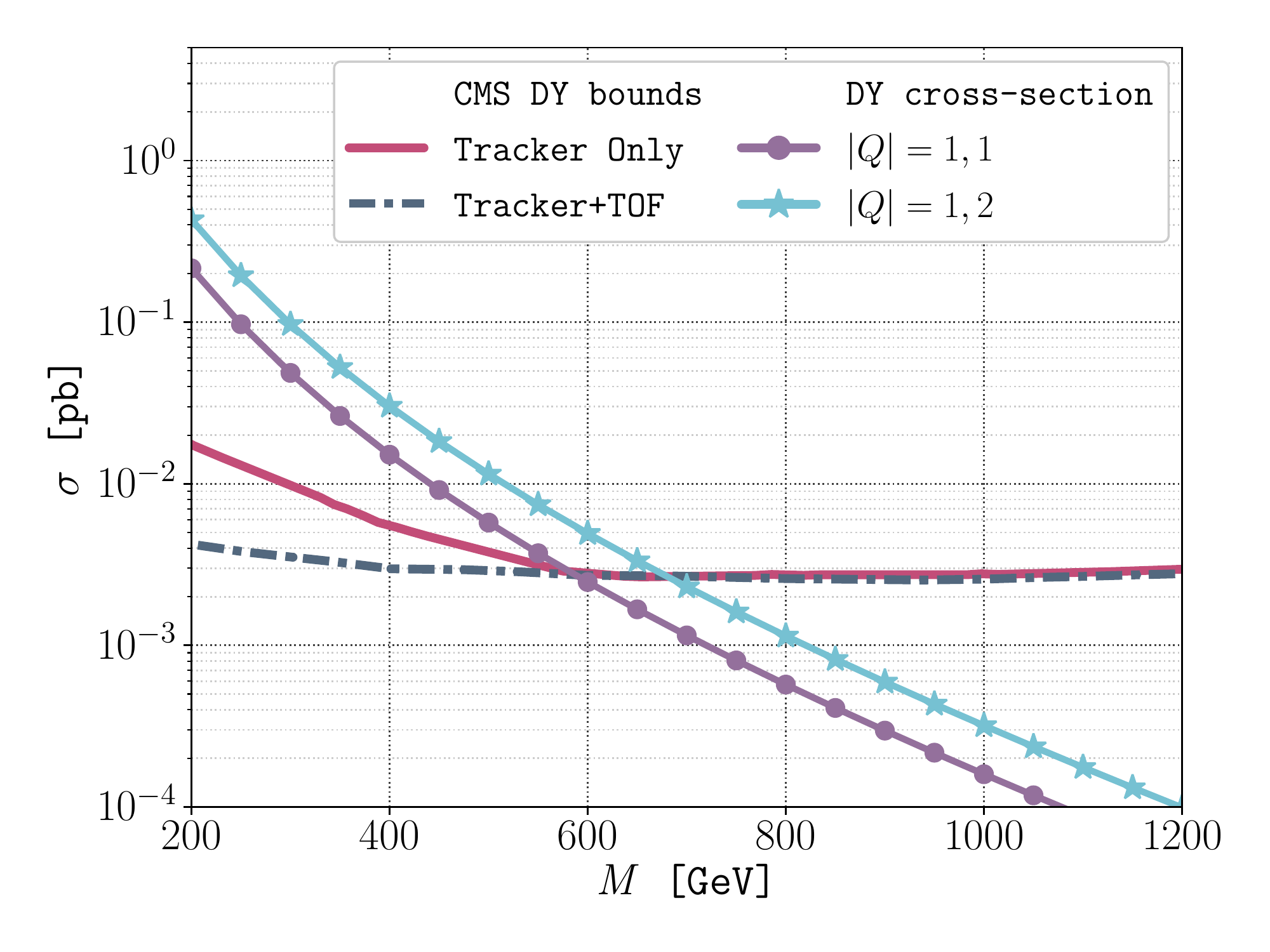}
	\hfill
	\includegraphics[trim= 10 10 20 10, clip, width=0.49\linewidth]{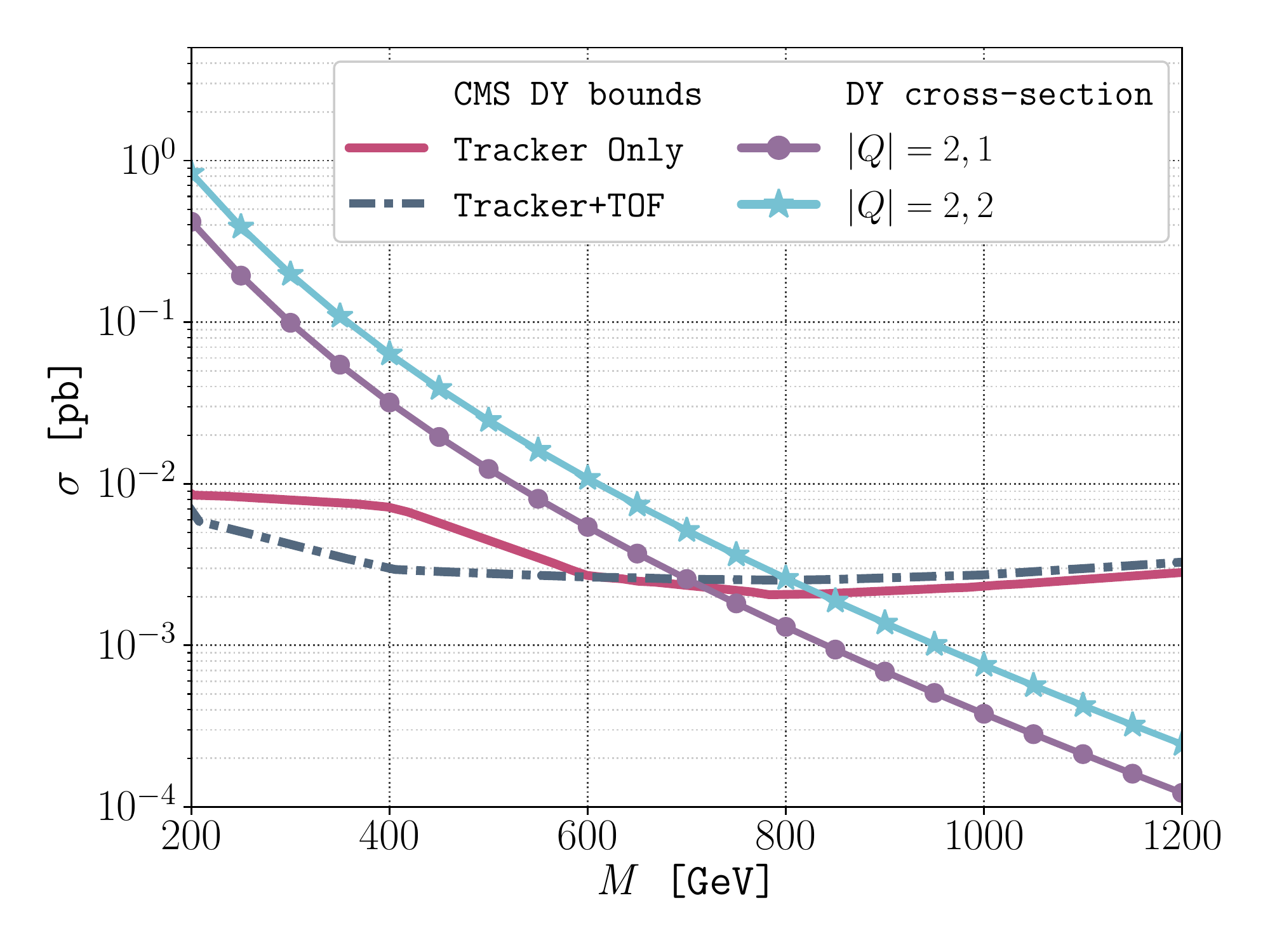}
\caption{\it Drell-Yann cross-section and experimental bounds for the exotic leptons at LHC (13 TeV) as a function of their mass $M$ for $|Q|=1$ (left panel) and $|Q|=2$ (right panel). The purple lines with circle shaped markers are the cross section for 1 exotic lepton with mass $M$ and the cyan lines with star-shaped markers are for 2 degenerate exotic leptons with the same $|Q|$. The cross-sections were computed at LO (see e.g.~\cite{DelNobile:2009st,DiLuzio:2015oha}) with MSTW2008 PDFs \cite{Martin:2009iq}. The curves without markers show the experimental results obtained by CMS \cite{Khachatryan:2016sfv} using only the tracker (continuous dark red line) and the tracker+TOF technique (dot dashed dark blue line).}
\label{fig:DYboundsLHC}
\end{figure}

The exotic leptons can be produced at hadron colliders through the Drell-Yann process. We show in Figure~\ref{fig:DYboundsLHC} the production cross sections and experimental limits \cite{Khachatryan:2016sfv} 
on the cross section at LHC for exotic leptons with $|Q|=1,\, 2$, which corresponds to $Y=3/2$. For the fully degenerate case, one must consider the production of 2 indistinguishable leptons with the same $|Q|$ and in that case, the bound on their mass is $M \gtrsim 840$ GeV, which comes from the states with $|Q|=2$. 

If one of the families is heavier than the other but each doublet is still degenerate, the bound relaxes to $M \gtrsim 720$ GeV. The situation of different masses in the doublet, unfavoured by EW precision data as explained before, would relax the bound for the $|Q|=1$ lepton in $~80$ GeV. Although the experimental results allow to further relax the bound for $|Q|=1$ in around $~40$ GeV, this would allow the decay of the $|Q|=2$ lepton via $W$ boson. We leave the phenomenological analysis of such case, together with the case of mixing among exotic lepton families, for future work.

Finally, if $Y=3/2$ (exactly), some exotic fermions can mix with SM leptons $\ell$
(unless other discrete symmetries are imposed). 
In such a case the phenomenology is rather different: $Q=2$ states can decay into a $W$ and a $Q=1$ fermion, 
while the latter can mix with SM leptons and decay into $Z\ell$ or $h\ell$. 
The lower bounds for these scenarios at LHC are around 500 GeV \cite{Ma:2014zda}.

\section{Conclusions}\label{conclusions}

In this paper we have investigated the structure of axion EFTs, 
focussing in particular on axion couplings to massive chiral gauge fields 
(most notably, SM electroweak gauge bosons)  
and their connection with UV models. 
Against naive expectations, Ref.~\cite{Quevillon:2019zrd} found that axion couplings to electroweak 
gauge bosons are not entirely captured by the mixed anomalies between the PQ and the gauge symmetries, 
even in the EFT limit when all the heavy fermionic degrees of freedom have been integrated out. 
This has to be compared, instead, with the more standard case of axion couplings to massless gauge fields 
(e.g.~photons and gluons) which are constrained by anomaly matching to always be linked to a UV PQ anomaly. 

In this work, we have provided a new understanding of this phenomenon.   
The key point is that massive gauge fields have longitudinal parts that can be used to write down non-anomalous gauge-invariant operators which 
contribute to the physical amplitudes but not to the anomalies. 
This is exemplified first in an abelian toy model, both within an EFT approach and an explicit matching between UV and IR dynamics upon 
integrating out heavy chiral fermions. In fact, the non-anomalous and non-decoupling terms are generated   
when the massive gauge fields are chiral with respect to some heavy fermions, 
which (dominantly) acquire their mass from the same source of symmetry breaking 
that gives mass to the chiral gauge fields. 

The generalization to the non-abelian case (relevant for the SM) 
brings in an important new feature: if axion couplings to vector bosons 
were solely determined by UV PQ anomalies, $SU(2)_L \times U(1)_Y$ invariance 
would induce correlations among the different EFT interactions. 
These can be expressed in terms of sum-rules for axion couplings to vector mass eigenstates 
(cf.~\eq{sumRuleEFTCoeffs}), which could be tested at low-energy within the EFT approach. 
On the other hand, 
non-anomalous gauge-invariant operators are generically expected if the electroweak 
symmetry is realized non-linearly 
and they allow to evade the above mentioned sum-rules for axion couplings to vector bosons. 
Hence, a violation of those sum-rules would clearly represent a smoking-gun signature 
for a UV-completion of the axion EFT which contains heavy fermions that are 
chiral with respect to the SM gauge group. 

We finally provided an explicit example of a phenomenologically relevant 
chiral extension of the SM, 
featuring an extended Higgs sector with an axion field whose couplings are induced by new heavy fermions 
whose mass is dominated by electroweak symmetry breaking vevs. 
In such a case, we verified that the general expectations 
discussed above
about axion couplings to 
electroweak gauge bosons are met in the EFT limit, 
such as e.g.~the breaking of the sum-rules for axion couplings to vector mass eigenstates. 
We further assessed the phenomenological viability of such a setup by 
inspecting electroweak precision tests, Higgs couplings and direct searches. 
As a side result, we showed that a certain class of chiral extensions of the SM 
(broadly understood as heavy fermions whose mass is dominantly due to electroweak symmetry breaking) 
is still viable, and thanks to an extended Higgs sector it could also hide from 
further scrutiny at the HL-LHC, although direct searches push it at the boundary of perturbativity.

\acknowledgments
We thank Thomas Biek\"otter, Emilian Dudas, Tony Gherghetta, Stefan Pokorski, J\'er\'emie Quevillon and Pablo Qu\'\i lez for useful discussions. 
This work is supported by the Deutsche Forschungsgemeinschaft under Germany's Excellence Strategy  EXC 2121 ``Quantum Universe'' - 390833306.
The work of C.G. and A.R. was also supported by the International Helmholtz-Weizmann Research School for Multimessenger Astronomy, largely funded through the Initiative and Networking Fund of the Helmholtz Association.
The work of L.D.L.~is supported by the Marie Sk\l{}odowska-Curie 
Individual Fellowship grant AXIONRUSH (GA 840791).

\appendix

\newcommand{\ii}{{\hat{\imath}}}
\newcommand{\jj}{{\hat{\jmath}}}

\section{EFT matching with a product of abelian gauge groups}\label{appendixMoreGaugeFields}

In this appendix, we provide explicit loop computations to match a model with a collection of heavy chiral fermions coupled to a product of abelian gauge bosons to an EFT Lagrangian of the type discussed in Section~\ref{physicalEFTsection}, involving axion terms as well as as GCS operators.\footnote{We followed a traditional approach with an explicit calculation of one-loop Feynman diagrams to match the UV-model onto the EFT Lagrangian. It would be interesting to re-derive the results more directly from the universal one-loop effective action, see Ref.~\cite{Cohen:2020fcu} and references therein for a recent review.} In this appendix, gauge couplings will be absorbed in the gauge fields, and they can be reinstated by making the replacement $A_{i,\mu}\rightarrow g_iA_{i,\mu}$ for each gauge field in all the formulae.

{Such computations have already been performed in the literature, for instance in Ref.~\cite{Anastasopoulos:2006cz} whose approach we follow closely (see also Ref.~\cite{Michaels:2020fzj} for a recent and similar computation relevant for radiative decays of the $Z$ boson). However, those discussions usually concern models of additional gauge symmetries, and to our knowledge rarely models of axions. Therefore, as in Section~\ref{physicalEFTsection}, we repeat the necessary details and insist on the treatment proper to axion models.}

\subsection{Explicit loop computation}\label{appendix:detailsIntegration}

We consider a heavy (chiral) fermion, $\psi$, coupled to several Abelian gauge fields, $A_{i,\mu}$. It will acquire its mass via a Yukawa interaction to a scalar field, $\phi$, whose imaginary component gives rise to an axion, $\theta$. The Lagrangian of the model is as follows:
\be
\cL_\psi = i\overline{\psi}\gamma^\mu\left(\partial_\mu-i[\alpha_i-\beta_i\gamma_5]A_{i,\mu}\right)\psi-y(\overline{\psi_{L}}\psi_{R}\phi+h.c.) \ ,
\label{lagHeavyFermionAppendix}
\ee
where $\psi_{R/L}=\frac{1\pm\gamma_5}{2}\psi$ (resp. $\phi$) have charges $q_{R/L}^i=\alpha_i\mp\beta_i$ (resp. $q^i_\phi=q^i_L-q^i_R=2 \beta_i$) with respect to the gauge field $A_i$, which again can be the ``fake'' PQ gauge field (henceforth, hatted indices $\hat{\imath}, \hat{\jmath}, \hat{k}$ will denote physical gauge symmetries, in particular not the PQ symmetry). 

As we explicitly show below, at energies below the mass of the heavy fermion, the interactions between the axions and the gauge fields can be matched to the following EFT Lagrangian,
\be
\cL_\text{EFT}\supset-\frac{3 \alpha_i \alpha_j + \beta_i \beta_j}{48\pi^2}\frac{\theta}{f}F_i\tilde{F_j} +\frac{E_{ij,k}}{8\pi^2}A_{i,\mu}A_{j,\nu}\tilde{F_k}^{\mu\nu}\ ,
\label{EFT:afterFermionAppendix}
\ee
where the sum over $i,j(,k)$ is implicit and we wrote $\phi=\frac{f}{\sqrt{2}}e^{i\frac{\theta}{f}}$. The GCS coefficients that are needed for our discussion are given by
\be
E_{\text{PQ}\,\ii, \jj}=2\left(\alpha_\text{PQ}\alpha_\ii+\frac{1}{3}\beta_\text{PQ}\beta_\ii\right)\beta_\jj
\ , \quad E_{\ii\jj,\text{PQ}}=2(\alpha_\ii \beta_\jj-\beta_\ii\alpha_\jj)\alpha_\text{PQ} 
\ .
\label{GCSwithOnePQAppendix}
\ee
Actually, only $E_{\text{PQ}\,\ii, \jj}$ is relevant for PQ anomaly matching since $E_{\ii\jj,\text{PQ}}$ does not contribute to the mixed PQ-gauge anomaly, but the value of the latter will be needed to check the invariance of the EFT Lagrangian under a physical gauge transformation.

\subsubsection{Axion terms}

\begin{figure}[!th]
\centering
\includegraphics[width=0.7\textwidth]{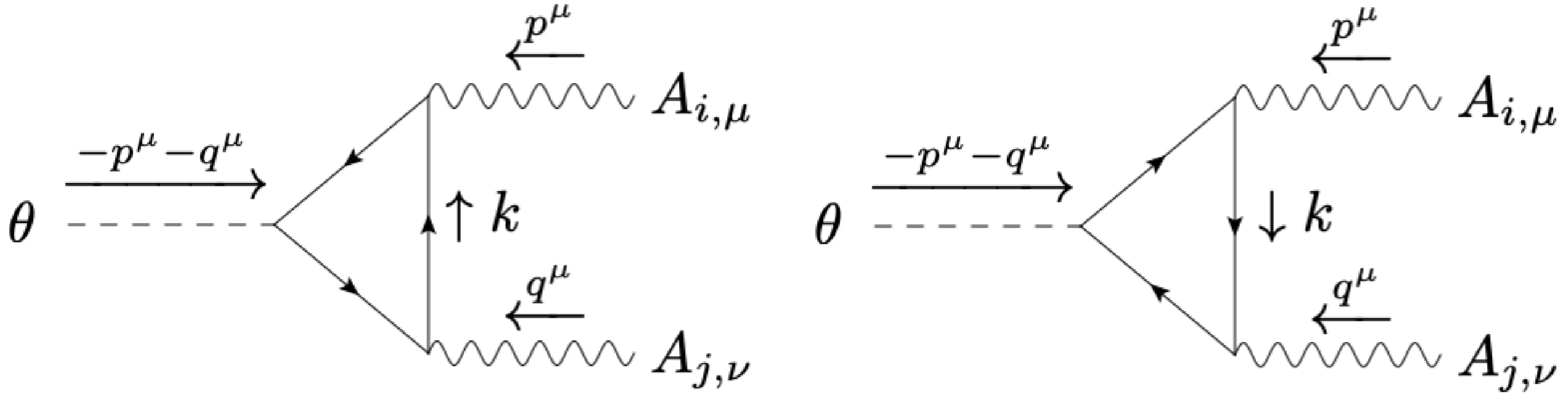}
\caption{\it One-loop contributions to the axion-gauge field coupling}
\label{axionDiags}
\end{figure}

The axion terms in the EFT arise via the (off-shell) diagrams of Fig.~\ref{axionDiags} (note that, when computing the one-loop EFT below the fermion masses, we should not compute diagrams with axions or gauge bosons propagators, since this would be a double counting with respect to amplitudes computed in the EFT).

Cutting off the external legs and matching with the Lagrangian coefficient $-c_{ij,\mu\nu}\frac{\theta}{v}A_{i,\mu}A_{j,\nu}$, we get
\begin{eqnarray}
c_{ij,\mu\nu}=&&\frac{m_\psi}{f}\int \frac{d^4k}{(2\pi)^4}\Tr(\gamma_5\frac{\cancel{k+p}+m_\psi}{(k+p)^2-m_\psi^2+i\epsilon}\Gamma^\mu_i\frac{\cancel{k}+m_\psi}{k^2-m_\psi^2+i\epsilon}\Gamma^\nu_j\frac{\cancel{k-q}+m_\psi}{(k-q)^2-m_\psi^2+i\epsilon})\nonumber\\
&&+(i,\mu,q\leftrightarrow j,\nu,q) \ ,
\label{axionDiagramsCalc}
\end{eqnarray}
where we defined $\Gamma^\mu_i\equiv\gamma^\mu(\alpha_i-\beta_i\gamma_5)$. To extract from this expression the piece proportional to the Levi--Civita tensor $\epsilon^{\mu\nu\rho\sigma}$, it is enough to focus on terms with one or three $\gamma_5$ in the trace. The corresponding expression is finite, and when evaluated in the $m_\psi\rightarrow\infty$ limit it corresponds to the first term written in \eqref{EFT:afterFermionAppendix}.

Note also that, if we kept the chiral fermion mass finite, the form factors in \eqref{axionDiagramsCalc} (and below in \eqref{GCSDiagramsCalc}) map to a tower of higher-dimensional operators in the EFT, see, e.g., Appendix~D of Ref.~\cite{Bonnefoy:2018ibr} for an explicit example.

\subsubsection{Generalised Chern--Simons terms}

\begin{figure}[!th]
\centering
\includegraphics[width=0.7\textwidth]{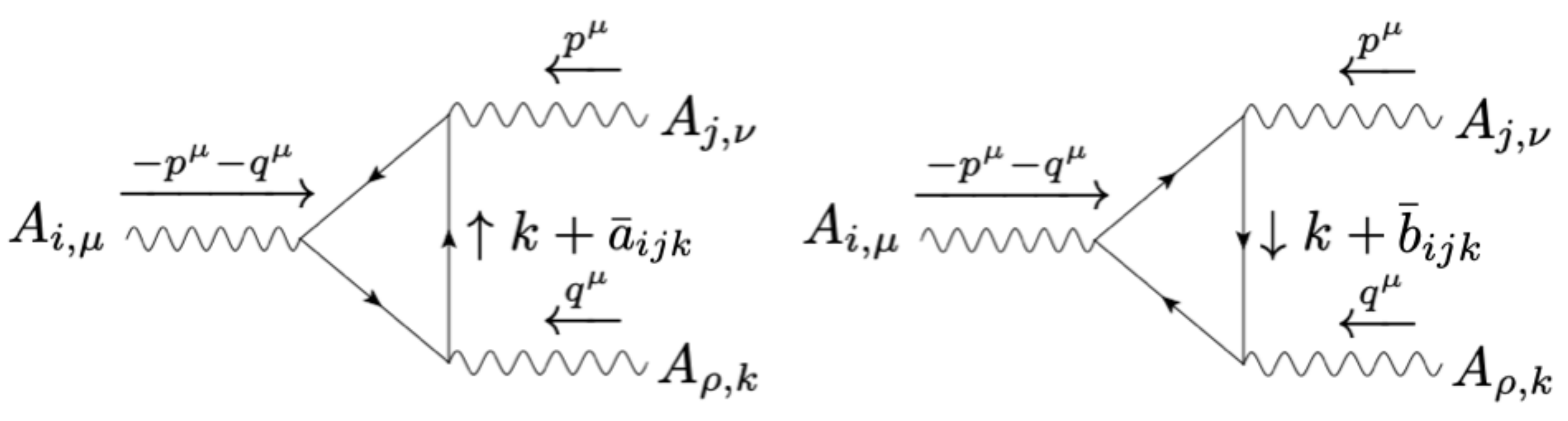}
\caption{\it One-loop contributions to the triple gauge field coupling}
\label{GCSDiags}
\end{figure}

To compute the GCS couplings, the relevant diagrams are that of Fig.~\ref{GCSDiags}.

They read
\be
\bead
&\int \frac{d^4k}{(2\pi)^4}\Tr(\frac{\cancel{k+a_{ijk}}+m_\psi}{(k+\bar{a}_{ijk})^2-m_\psi^2+i\epsilon}\Gamma^\rho_k\frac{\cancel{k+\bar{a}_{ijk}-q}+m_\psi}{(k+\bar{a}_{ijk}-q)^2-m_\psi^2+i\epsilon}\Gamma^\mu_i\frac{\cancel{k+\bar{a}_{ijk}+p}+m_\psi}{(k+\bar{a}_{ijk}+p)^2-m_\psi^2+i\epsilon}\Gamma^\nu_j)\\
&+(j,\nu,p,\bar{a}_{ijk}\leftrightarrow k,\rho,q,\bar{b}_{ijk}) \ .
\label{GCSDiagramsCalc}
\eead
\ee
We kept track of the momentum routing ambiguity by introducing two shift vectors $\bar{a}_{ijk},\bar{b}_{ijk}$ in the computation since, although the GCS are associated to finite terms, each diagram will contribute via a linearly divergent expression and shifts of $k$ in one diagram have consequences.
For the axion couplings, each of the two diagrams in \eqref{axionDiagramsCalc} is at most logarithmically divergent (due to the Dirac trace, there is no $k^3$ term in the numerator), so there was no need to introduce shift vectors.

To extract the GCS terms, we focus again on the terms with odd numbers of $\gamma_5$ in the trace. Furthermore, we only care about non-decoupling EFT terms in the $m_\psi\rightarrow\infty$ limit, which must be of the form $A^3$ or $\partial A A^2$ since any higher dimensional operator has to be suppressed by $m_\psi$, the only large scale which enters in the computation. Thus, it is enough to look at the zeroth and first order of \eqref{GCSDiagramsCalc} in $p,q,a,b$. The end result is
\be
\bead
-\frac{\epsilon^{\mu\nu\rho\sigma}}{4\pi^2}\bigg(p_\sigma\left[\alpha_j\alpha_k\beta_i+\alpha_i\alpha_k\beta_j-\alpha_i\alpha_j\beta_k+\frac{\beta_i\beta_j\beta_k}{3}\right]-q_\sigma\left[\alpha_j\alpha_k\beta_i-\alpha_i\alpha_k\beta_j+\alpha_i\alpha_j\beta_k+\frac{\beta_i\beta_j\beta_k}{3}\right]&\\
+\frac{(\bar{b}_{ijk}-\bar{a}_{ijk})_\sigma}{2}\left[\alpha_j\alpha_k\beta_i+\alpha_i\alpha_k\beta_j+\alpha_i\alpha_j\beta_k+\beta_i\beta_j\beta_k\right]&\bigg) \ .
\eead
\label{GCSLoopAmplitude}
\ee
To be consistent with anomaly computations, see e.g. Ref.~\cite{Weinberg:1996kr}), we must relate the two shift vectors as
\be
-\bar{b}_{ijk}=\bar{a}_{ijk}=a^{(p)}_{ijk}p+a^{(q)}_{ijk}q \ .
\ee
This allows one to match \eqref{GCSLoopAmplitude} to the amplitude obtained from the GCS term ${E_{ij,k}}A_{i}A_{j}\tilde{F_k}$, of the effective Lagrangian \eqref{EFT:afterFermionAppendix}, which in momentum space reads
\be
-\frac{\epsilon^{\mu\nu\rho\sigma}}{4\pi^2}((E_{jk,i}-E_{kj,i})(-p-q)_\sigma+(E_{ki,j}-E_{ik,j})p_\sigma+(E_{ij,k}-E_{ji,k})p_\sigma) \ .
\label{GCSamplitude}
\ee
The GCS couplings $E_{ij,k}$ are then uniquely determined in terms of the charges $\alpha_i, \beta_i$ and the shift vectors $a_{ijk^{(p)}}$ by requiring that they obey the following two conditions: 
\be
\text{(i) } \ E_{ij,k}=-E_{ji,k} \ ,\quad \text{(ii) } \ E_{ij,k}+E_{jk,i}+E_{ki,j}=0 \ .
\ee
The first relation follows from the symmetric property under the exchange of $i$ and $j$ of the quantity  $A_i A_j \tilde{F}_k$, while the second relation is a consequence of the fact that $A_i A_j \tilde{F}_k + A_j A_k \tilde{F}_i + A_k A_i \tilde{F}_j$ is a total derivative. 
The two expressions  \eqref{GCSLoopAmplitude} and \eqref{GCSamplitude} will then be equivalent for arbitrary momenta $p$ and $q$ and arbitrary charges $\alpha_i$ and $\beta_i$ provided that the  shift vectors satisfy:
\be
a_{ijk}\equiv a_{ijk}^{(p)}=-a_{kij}^{(q)} \ , \quad a_{ijk}=a_{jik} \ , \quad a_{ijk}+a_{kij}+a_{jki}=1 \ .
\label{shiftVectorConstraints}
\ee
It then follows that
\be
E_{ij,k}=(a_{ikj}-a_{jki})\left(\alpha_i\alpha_j+\frac{1}{3}\beta_i\beta_j\right)\beta_k+(1-a_{ijk})(\alpha_i\beta_j-\beta_i\alpha_j)\alpha_k \ .
\label{EFT:afterFermionGCS}
\ee
The shift vectors are chosen in order to enforce that the physical gauge symmetries are conserved, namely that the EFT is invariant under the physical gauge symmetries.
Let us consider the following transformation of the fields,
\be
\psi_{L/R}\rightarrow e^{i\epsilon_iq_{L/R}^i}\psi_{L/R} \ , \quad \theta \rightarrow \theta + \epsilon_i(q_L^i-q_R^i)f \ , 
\quad A_{i,\mu}\rightarrow A_{i,\mu}+\delta_{ij}\partial_\mu\epsilon_j \ .
\ee
Using the expression \eqref{EFT:afterFermionGCS} of the GCS coefficients $E_{ij,k}$ in terms of the shift vectors, the variation of the EFT Lagragian \eqref{EFT:afterFermionAppendix} reads\footnote{Our results are a factor of $2$ off with respect to the ones in \cite{Anastasopoulos:2006cz}, so that anomaly cancellation holds in the low-energy EFT in our case. It can be verified by an explicit computation in the UV model that the RHS of \eqref{anomalyVariation} corresponds to the variation of the quantum effective action associated to the heavy fermion, see for instance Ref.~\cite{Weinberg:1996kr}. This holds irrespective of the fact that $i$ corresponds to a genuine gauge symmetry or the PQ one.}
\be
\delta_i \cL_\text{EFT}=-\frac{(1-a_{jki})D^{ijk}}{32\pi^2}\epsilon_iF_j\tilde{F}_k \ ,
\label{anomalyVariation}
\ee
where $D^{ijk}\equiv q_L^iq_L^jq_L^k-q_R^iq_R^jq_R^k = 2 \beta_i\beta_j\beta_k + 2 (\alpha_i\alpha_j\beta_k+\alpha_j\alpha_k\beta_i+\alpha_k\alpha_i\beta_j)$ is nothing else but 
the $U(1)_i\times U(1)_j\times U(1)_k$ anomaly polynomial.
By considering $i$ and $j$ to be gauge symmetries and $k$ the PQ symmetry, which has possible non-vanishing mixed gauge anomalies, the consistency of the EFT Lagrangian at the quantum level, namely $\delta_\ii \cL_\text{EFT}=0$,  forces to choose
\be
a_{\text{PQ} \hat{\imath}\hat{\jmath}}=1 \ , \quad a_{\hat{\imath}\hat{\jmath}\text{PQ}}=-1 \ ,
\ee
where the last equality follows from \eqref{shiftVectorConstraints}. Plugging back these values of the shift vectors, we arrive at the expression of the GCS coefficients announced in \eqref{GCSwithOnePQAppendix}.

It can be checked in particular that these results guarantee that the whole EFT Lagrangian remains invariant under the action of an unbroken symmetry, i.e., a symmetry under which the scalar field $\phi$ is neutral, $q^\ii_\phi=\beta_\ii=0$. In that case the axion $\theta$ itself does not shift under the action of $A_\ii$. The individual GCS terms do, but in a correlated way. Indeed,  when $\beta_\ii=0$, from \eqref{GCSwithOnePQAppendix}, we obtain
\bes
\cL_\text{EFT}\supset \frac{\alpha_\text{PQ}\alpha_{\hat{\imath}}\beta_{\hat{\jmath}}}{2\pi^2}\left(A_{\text{PQ},\mu}A_{{\hat{\imath}},\nu}\tilde{F}_{\hat{\jmath}}^{\mu\nu}+A_{{\hat{\imath}},\mu}A_{{\hat{\jmath}},\nu}\tilde{F}_\text{PQ}^{\mu\nu}
\right)= -\frac{\alpha_\text{PQ}\alpha_{\hat{\imath}}\beta_{\hat{\jmath}}}{2\pi^2}A_{\text{PQ},\mu}A_{{\hat{\jmath}},\nu}\tilde{F}_{\hat{\imath}}^{\mu\nu} +\text{tot. derivative}\ ,
\ees
that is  gauge invariant with respect to $A_\ii$, as it should.

\subsection{Anomaly matching with axion terms only}\label{appendix:prescription}

In the main text, we argued that, contrary to the simple case of the axion decay into two photons, in general the phenomenology of an axion coupled to gauge fields  is not uniquely specified by the knowledge of mixed gauge-PQ UV anomalies as, indeed, the latter are only reproduced when combining the axion and the GCS terms. Still, it is interesting to ask how particular is the case of photons and when it can be generalised, i.e., what are the conditions for the axion terms alone to reproduce the UV anomalies?\footnote{A connected case is the one of a single physical massive gauge field, the axion $\theta$ becoming the longitudinal component.
The contribution of a heavy fermion of the $U(1)^3$ gauge anomaly is then fully captured, in generic gauge, by the axion term --  it is not possible to write a GCS term involving a single gauge field, see the anomaly inflow on a local string, Ref.~\cite{Callan:1984sa}. This is consistent with the well-known result of Ref.~\cite{Preskill:1990fr} that spontaneously broken gauge symmetry with anomalous fermion content can be consistently quantized.}

When the heavy fermion is coupled to massless gauge bosons only, like the photons, the scalar field $\phi$ has to be gauge neutral, $\beta_\ii=\beta_\jj=0$ and, according to \eqref{GCSwithOnePQAppendix}, the GCS terms are absent. So the axion term is the only one that can reproduce the anomalous shift under a PQ transformation.
However, when the heavy fermion has also an axial coupling to at least one (massive) gauge field, $\beta_\ii\not = 0$, then a GCS term is needed in the EFT to add up to the shift of the axion term in order to reproduce the full UV anomaly.

But even in the case of an axial gauge symmetry, it  might still be possible that the mixed PQ anomaly is borne by the  axion term only provided that the PQ symmetry is conveniently chosen.  In these models, there need to unbroken vector-like symmetries that can be used to redefine what one calls the PQ symmetry. For instance, the minimal model in \eqref{lagHeavyFermionAppendix} has a ``$\psi$-number'' symmetry under which $\psi\rightarrow e^{i\epsilon_\psi}\psi$. Similarly, the model in Section~\ref{eq:chiralextSM} has two unbroken lepton number symmetries. The PQ charges are defined up to these vector-like transformations. They do not affect the axion couplings~\cite{Quevillon:2019zrd,Quevillon:2020hmx}, but they modify the GCS terms and the UV anomalies, in a way consistent with anomaly matching as discussed above. For the  axion term to capture the full PQ anomaly, we need to impose that
\be
\delta_\text{PQ}\left(\frac{E_{\text{PQ}\ii,\jj}}{8\pi^2}A_{\text{PQ},\mu}A_{\ii,\nu}\tilde{F}_\jj^{\mu\nu}\right)=-\frac{E_{\text{PQ}\ii,\jj}}{16\pi^2}\epsilon_{\text{PQ}}F_\ii \tilde{F}_\jj=0 \ .
\ee
It is therefore necessary and sufficient that $E_{\text{PQ} {\hat{\imath}}, {\hat{\jmath}}}+E_{\text{PQ}{\hat{\jmath}},{\hat{\imath}}}=0$, i.e., given the explicit expression \eqref{GCSwithOnePQAppendix} of the GCS coefficients, 
\be
3(\alpha_\ii\beta_\jj+\beta_\ii\alpha_\jj)\alpha_\text{PQ}+2\beta_\ii\beta_\jj\beta_\text{PQ}=0\ .
\label{eq:goodPQ}
\ee
Under such a condition, the mixed PQ gauge anomaly coefficient simply becomes
\be
D^{\text{PQ}\ii\jj}=\frac{(3 \alpha_\ii\alpha_\jj + \beta_\ii\beta_\jj)}{3}q^\text{PQ}_\phi \ ,
\label{prescriptionAppendix}
\ee
where $q^\text{PQ}_\phi=q_L-q_R=2 \beta_{\text{PQ}}$ is the PQ charge of the scalar field $\phi$. We recognize here the coefficient of the axion term (times the axion charge). Equation~(\ref{eq:goodPQ}) generalizes the prescription in \eqref{prescriptionFormula} derived for a model with a single gauge symmetry. {The integration of several fermions at once, in particular when they form an anomaly-free set with respect to the gauge symmetries, is similar to Eq.~\eqref{EFTwithNoGaugeAnomaly} (see also Appendix~\ref{PQMatchingExampleAppendix}).}

Note that in the presence of several chiral gauge fields and for generic values of the fermion gauge charges, the axion couplings are not expected to match all mixed anomalies of the PQ symmetry at once, whatever the choice of the PQ charges of the UV fermions. In Section~\ref{eq:chiralextSM}, we showed that this fact has observational consequences for non-abelian theories, which take the form of the violation of sum rules. For abelian theories, the statement is mostly formal, but it still applies, as can be seen in simple models such as the one of \eqref{lagHeavyFermionComplete}, where no \sout{UV PQ anomaly coefficient} {PQ charge assignment in the UV can generate a set of anomaly coefficients which} captures all axion couplings at once. {An illustration of this is obtained when restricting the gauge theory of the model in section~\ref{eq:chiralextSM} to the photon and the $Z$ boson only (or equivalently, to the hypercharge and $T_3$ generators in the UV). It is then shown in appendix~\ref{PQMatchingExampleAppendix} that the PQ charge assignment in the UV cannot match the anomaly coefficients and the couplings $a\gamma\gamma,a\gamma Z,aZZ$ in the EFT.} However, for each individual axion coupling, there exists one convenient PQ charge obtained from 
\eqref{eq:goodPQ}. Note that if the heavy fermion has a vector-like charge under one of the gauge symmetries and a chiral one under the other (e.g. $\beta_\ii=0,\beta_\jj\neq0$), then the suitable PQ charge is purely axial ($\alpha_\text{PQ}=0$).

\subsection{CCWZ approach to the EFT}\label{CCWZsection}

In order to connect to formulae such as \eqref{operator1Quator}-\eqref{EFTarranged}, we would like to point out that it is always possible to rearrange the EFT Lagrangian \eqref{EFT:afterFermionAppendix}, and more generally any EFT of axions $a_n$ and gauged Goldstone bosons $\theta_\ii$, as a combination of the following terms,
\be
-\frac{\cC_{n\hat  i\hat j}}{16\pi^2f_n}a_nF_{\ii,\mu\nu}\tilde{F}_\jj^{\mu\nu} \ ,\quad -{\cal E}_{n\ii\jj}\bigg(\frac{\partial_\mu a_n}{f_n}-A_{n,\mu}^\text{PQ}\bigg)\bigg(\frac{\partial_\nu\theta_\ii}{m_{A_\ii}}-A_{\ii,\nu}\bigg)\, \tilde F^{\mu\nu}_\jj  \ .
\label{anomAndNonAnomTerms}
\ee
(plus other possible PQ- and gauge-invariant terms which do not involve the axions), where $A_\jj$ can be massless but $A_\ii$ has to be massive. The two kinds of terms are respectively anomalous and gauge-invariant, generalizing \eqref{operator1} and \eqref{operator1Quator}. The several axions $a_n$ and gauge GBs $\theta_\ii$ find their UV origin in the phases of Higgs fields, and both operators are obtained by integrating out fermions. In order to fulfil PQ anomaly matching, we must therefore have $\cC_{n\ii\jj}=D^{\text{PQ}_n\ii\jj}$, the UV mixed anomaly coefficients of the $n$-th PQ symmetry. This rearrangement is straightforwardly done from our explicit formulae, but it can also be simply understood from a field redefinition of the UV theory. For that, we generalize further the UV models \eqref{lagHeavyFermionAppendix} and consider the case of a renormalizable UV theory involving several charged chiral fermions $\psi_u$ getting their masses via Yukawa couplings to several Higgs fields $\phi_X$:
\be
\cL_\text{UV} = i\overline{\psi}_u \gamma^\mu\left(\partial_\mu-i[\alpha_{i,u}-\beta_{i,u}\gamma_5]A_{i,\mu}\right)\psi_u-(y^X_{uv}\,\overline{\psi}_{u,L}  \psi_{v,R}\,\phi_X+h.c.) \ .
\label{lagHeavyFermionMoreGeneral}
\ee
The gauge fields are split in two categories: the ones with respect to which the fermions are vector-like and which can be massless (if they get a mass from some other source, it is irrelevant for our argument), and the necessarily massive ones under which the fermions are chiral. For those massive vectors $A_i$ (among which the PQ gauge field), it is possible to define fields $\theta_i$ out of the phases $\theta_X$ of the Higgs fields, that shift as $\theta_i\rightarrow m_{A_i}\delta^{j}_i\epsilon_j$ under the $j$-th gauge symmetry\footnote{Precisely, the mass term for the gauge fields,
\be
\cL\supset \frac{1}{2}M^2_{ij}A_{i,\mu}A_j^\mu \ ,
\ee
arises from the axion kinetic terms (themselves obtained from the Higgs fields kinetic terms)
\be
\cL\supset \frac{1}{2}(\partial_\mu\theta_I-q_I^iv_IA_{i,\mu})^2
\ee
so that $M^2_{ij}=q_I^iq_I^jv_I^2$. Diagonalizing $M^2=O^TM'^2O$, for an orthogonal matrix $O$ and $M'^2=\text{diag}(m_{A_i}^2)$, $A'_i=(OA)_i$ define the massive vector fields of masses $m_{A_i}$ and the associated gauge parameters are $\epsilon'_i=(O\epsilon)_i$. Thus, defining $X_{iI}=\frac{O_{ij}q_I^jv_I}{m_{A_i}}$ which is orthogonal (recall that we introduce fake gauge fields for each PQ symmetry, so in particular there is one massive field for each phase of a Higgs scalar), $\theta'_i=(X\theta)_i$ shifts as $m_{A_i}\delta^j_i\epsilon'_j$ under the gauge symmetries defined by the vectors $A'_i$. We can also define $q'_I{}^i=(O^Tq_I)^i$ and check that we consistently have $\frac{\theta_I}{v_I}=q'_I{}^i\frac{\theta'_i}{m_{A_i}}$. We work with the primed fields and drop the primes.}. Thus, we can make the fermions uncharged under the massive gauge fields by redefining them as follows \cite{Coleman:1969sm,Callan:1969sn},
\be
\psi_{u,L/R} \rightarrow e^{i\frac{\theta_i}{m_{A_i}}q^i_{u,L/R}} \psi_{u,L/R} \ .
\ee
Since the fermions are now uncharged (except under the vector-like gauge symmetries), there cannot remain any axion in the Yukawa couplings. Also, the field redefinitions modify the minimal coupling to gauge fields in the covariant derivative as follows:
\be
\overline{\psi_{u}}\gamma^\mu\left(\partial_\mu-i[\alpha_{i,u}-\beta_{i,u}\gamma_5]A_{i,\mu}\right)\psi_{u} \rightarrow \overline{\psi_{u}}\gamma^\mu\left(\partial_\mu+i[\alpha_{i,u}-\beta_{i,u}\gamma_5]\left[\frac{\partial_\mu\theta_i}{m_{A_i}}-A_{i,\mu}\right]\right)\psi_{u} \ ,
\ee
which is correctly gauge invariant when the fermions are neutral. Thus, integrating the latter out (in perturbation theory or beyond) cannot generate anything else than terms which are made out of $\frac{\partial_\mu\theta_i}{m_{A_i}}-A_{i,\mu}$ \cite{Preskill:1990fr,Feruglio:1992fp,Masiero:1992wd}, among which terms such as the second ones in \eqref{anomAndNonAnomTerms}. To such terms, one needs to add the jacobian contribution due to the anomalous transformation of the path integral measure \cite{Fujikawa:1980eg}, which are nothing but the first terms in \eqref{anomAndNonAnomTerms} for the physical axions $\theta_i\equiv a_n$, see the previous section for details. Anomaly matching between the UV and the IR is obvious in this approach.
\section{Axion couplings and PQ anomalies in SM chiral extensions}\label{appendixPheno}

In this appendix, we discuss the axion couplings in the SM chiral extension of Section \ref{eq:chiralextSM}, and we compare them to the UV PQ anomalies. In particular, we match them one by one using the prescription of \eqref{prescriptionAppendix}, which determines how to fix the fermion PQ charges so that the axion coupling under study is reproduced by the corresponding PQ anomaly. We show how this prescription should be properly used when the PQ symmetry is restricted by the fact that it should commute with gauge symmetries.

\subsection{General 2HDM assignment and axion couplings}\label{generalAssignmentAppendix}

For generality, we first present what the axion couplings would be if we assigned the two Higgses differently than in \eqref{eq:Yuk2H}. Let us denote $H_\psi$ the Higgs that appears in $\psi$'s Yukawa term (only one can appear if the PQ is to be exact and the axion classically massless). Our Yukawa sector thus looks as follows,
\beq
- \mathcal{L}_Y = 
y_{E_1} \bar L_1 E_1 H_{E_1} + y_{E_2} \bar L_2 E_2 H_{E_2} +
y_{N_1} \bar L_1 N_1 H_{N_1} + y_{N_2} \bar L_2 N_2 H_{N_2} + \text{h.c.} \, ,
\label{generalAssignment}
\eeq
and the axion content of each Higgs field, which can be $H_i$ or $\tilde H_i$, is written $H_\psi=e^{iX_{H_\psi}\frac{a}{f}}\left(0 \quad \frac{v_{H_\psi}}{\sqrt{2}}\right)^T$, so that the (Yukawa) axion fermion coupling is $\frac{y_\psi v_{H_\psi}}{\sqrt{2}}\bar\psi_L\psi_Re^{iX_{H_\psi}\frac{a}{f}}+h.c.$. In terms of the PQ charges of the fermions, one has 
\be
X_{H_{E_i}}=X_{L_i}-X_{E_i} \ ,\quad X_{H_{N_i}}=X_{L_i}-X_{N_i} \ .
\ee
We derive the axion EFT below the mass of the new fermions, using the formulae in Appendix \ref{appendixMoreGaugeFields}. For that, we define the massive eigenstates $N_i=N_i+N_{L_i}$, $E_i=E_i+E_{L_i}$, where
\be
L_1=\bmat N_{L_1}\\E_{L_1}\emat \ , \quad L_2=\bmat E_{L_2}\\N_{L_2}\emat \ ,
\ee
and extract their gauge couplings from
\begin{eqnarray}
\cL\supset&&\overline{L_i}\gamma^\mu\left(g\frac{\sigma^a}{2}W^a_\mu+g'(-1)^{i+1}YB_\mu\right)L_i+g'\overline{N_i}\gamma^\mu(-1)^{i+1}\left(Y+\frac{1}{2}\right)B_\mu N_i\nonumber\\
&&+g'\overline{E_i}\gamma^\mu(-1)^{i+1}\left(Y-\frac{1}{2}\right)B_\mu E_i \ .
\end{eqnarray}
Assuming equal masses within a $SU(2)$ doublet for simplicity ($m_{N_i}=m_{E_i}$), the axion couplings in the EFT read\footnote{With our assumption of equal masses within a doublet, the diagrams for the $aW^a\tilde W^b$ process can be obtained using our abelian formulae with $\alpha=\beta=\frac{1}{4}$, and supplemented by a factor 
\bes
\sum_{i=1,2 \text{(flavors)}}\sum_{j=1,2\text{($\psi^j$=N,E)}}(\sigma^{\{a}\sigma^{b\}})_{[1+i+j],[1+i+j]}X_{\psi^j_i} \ ,
\ees
where we defined $x_2=x\mod 2$, $[x]=x_2+2(1-x_2)$.}
\be
\cL\supset-g'^2\frac{(1+12Y^2)X^++6YX^-}{192\pi^2}\frac{a}{f}B\tilde B - g^2\frac{X^+}{192\pi^2}\frac{a}{f}W^a\tilde W^a\big\vert_\text{lin.}- gg'\frac{X^++6YX^-}{192\pi^2}\frac{a}{f}B\tilde W^3\big\vert_\text{lin.}
\label{axionCouplingsAppendix}
\ee
where 
\be
X^+=\sum_i\left(X_{H_{N_i}}+X_{H_{E_i}}\right) \ , \quad X^-=\sum_i\left(X_{H_{N_i}}-X_{H_{E_i}}\right) \ .
\ee
From this formula, we can immediately check that there are no (dimension 5) axion-gauge bosons couplings if all the Higgses in \eqref{generalAssignment} are expressed in terms of a single one, as given in \eqref{eq:Yuk1H}. Indeed, one then has $X_{H_{N_1}}+X_{H_{N_2}}=X_{H_{E_1}}+X_{H_{E_2}}=0$, such that $X^+=X^-=0$. Actually, this has a nice interpretation in terms of our discussion of section \ref{physicalEFTsection}. Let us focus on the $aB\tilde B$ coupling for simplicity. The mass terms and the axion couplings in \eqref{generalAssignment} have a $U(1)^5$ symmetry - broken to $U(1)^3$ by weak interactions -, spanned by the hypercharge $U(1)_Y$ and four fermion numbers $U(1)_{\psi=E_i,N_i}$. Thus the PQ symmetry must be a linear combination of those. We can choose it to be fully aligned with $U(1)_Y$ without affecting the discussion of anomaly matching in the axion EFT, since the axion does not shift under the vector-like fermion symmetries. Then, gauging again the PQ symmetry, one finds that the fields cannot differentiate between the PQ or the $B$ gauge field, since each fermion $\psi$ couples to a single combination of them, $Y_\psi(A_{\text{PQ},\mu}+A_\mu)$, hence there cannot be any GCS term in the EFT, simply because one cannot write a non-vanishing GCS term with a single gauge field. Also, each fermion couples to a single axion, the phase of the single Higgs, so that there is a single axion term. Consequently, the axion term is given by the $U(1)_Y^3$ anomaly coefficient, or by the $U(1)_\text{PQ}U(1)_Y^2$ anomaly coefficient, which are identical by assumption. Eventually, since the full set of heavy fermions has no $U(1)_Y^3$ anomaly, there is no possible axion coupling in the EFT.

\subsection{Matching with the UV PQ anomaly coefficients}\label{PQMatchingExampleAppendix}

We now verify that the prescription in \eqref{prescriptionAppendix} reproduces the axion couplings of \eqref{axionCouplingsAppendix}. Let us recall why such a prescription is needed. The interactions in \eqref{generalAssignment} have two unbroken lepton number symmetries $U(1)_{L_1}$ and $U(1)_{L_2}$, which are anomalous with respect to the chiral gauge symmetries $SU(2)_L \times U(1)_Y$. Therefore, the PQ UV anomalies can be modified by adding to the fermion PQ charges a component along those fermion numbers, namely by redefining $U(1)_\text{PQ}\rightarrow U(1)_\text{PQ}+\alpha\left(U(1)_{L_1}+U(1)_{L_2}\right)$, with $\alpha$ an arbitrary number (the antisymmetric combination $U(1)_{L_1}-U(1)_{L_2}$ is anomaly free). This number can be used to modify each mixed PQ anomaly so that it matches the corresponding axion coupling in \eqref{axionCouplingsAppendix}. We show how this is achieved in what follows.  

For that, we compute the anomalous terms arising after a PQ rotation that removes the axion from the mass terms,
\be
L_i \to e^{i X_{L_i}\frac{a}{f}} L_i \, , \quad N_i  \to e^{i X_{N_i}\frac{a}{f}} N_i \, , \quad E_i \to e^{i X_{E_i}\frac{a}{f}} E_i \, ,
\ee
where $i=1,2$. The anomalous terms read
\be
\delta\cL=-g^2\frac{\cA_W^{ab}}{16\pi^2}\frac{a}{f}W^a\tilde W^b-g'^2\frac{\cA_B}{16\pi^2}\frac{a}{f}B\tilde B \ ,
\label{anomalousTermsLagrangianAppendix}
\ee
where the anomaly polynomials are
\be
\bead
\cA_W^{ab}= \frac{(X_{L_1}+X_{L_2})\delta^{ab}}{2} \ , \quad \cA_B=\ \frac{X^+-2(X_{L_1}+X_{L_2})}{4}+YX^-+Y^2X^+ \ ,
\eead
\label{anomalousTermsAppendix}
\ee
where $X^\pm$ have been defined in appendix \ref{generalAssignmentAppendix}, and we recognize in particular the contribution of the unbroken anomalous fermion number $U(1)_{L_1}+U(1)_{L_2}$.

We now specialize to the Higgs assignment in \eqref{eq:Yuk2H}, for which we have the relations \eqref{PQrelations}, which imply $X^+=2(X_1-X_2),X^-=0$ so that we get
\be
\cA_W^{ab}=\frac{(X_{L_1}+X_{L_2})\delta^{ab}}{2} \quad
\cA_B=-\frac{X_{L_1}+X_{L_2}+(X_2-X_1)(1+4Y^2)}{2} \ .
\ee
Restricting \eqref{anomalousTermsLagrangianAppendix} to the neutral mass eigenstates, one finds
\be
\bead
-\frac{16\pi^2}{e^2}\delta\cL=\ &(X_1-X_2)\left(2Y^2+\frac{1}{2}\right)F\tilde F\\
&+\left(-t_W\left[(X_1-X_2)\left(4Y^2+1\right)-(X_{L_1}+X_{L_2})\right]+t^{-1}_W[X_{L_1}+X_{L_2}]\right)F\tilde Z\\
&+\left(t_W^2\left[(X_1-X_2)\left(2Y^2+\frac{1}{2}\right)-\frac{X_{L_1}+X_{L_2}}{2}\right]+t^{-2}_W\frac{X_{L_1}+X_{L_2}}{2}\right)Z\tilde Z \ .
\eead
\label{PQanomaliesAppendix}
\ee
Note that \eqref{PQanomaliesAppendix} verifies the constraints \eqref{constraintRelationCoefficients}, as it should. As expected the photon terms match with the ones in \eqref{axionPhotonZcouplings}. To match the photon-$Z$ terms, one needs an axial PQ, as discussed around \eqref{prescriptionAppendix}. However, given the relations \eqref{PQrelations}, it is impossible to define the PQ symmetry such that it is chiral on each heavy fermion, namely one cannot enforce
\be
X_{L_1} + X_{E_1} = 0 \, , \qquad  X_{L_1} + X_{N_1}= 0 \ , \quad X_{L_2} + X_{E_2}= 0 \, , \qquad  X_{L_2} + X_{N_2}= 0 \, ,
\ee
unless the PQ charges of the Higgses are non generic, namely if they verify $X_1=-X_2$. This clash comes from the fact that we defined the PQ symmetry on the UV fields so that it commutes with all gauge symmetries. In particular, both components of the $SU(2)$ doublets $L_i$ have the same PQ charge. Starting from a generic $U(1)_\text{PQ}$ charge assignment, $X_1=-X_2$ can be reached by considering a suitable linear combination of the hypercharge and the original PQ symmetry. If we want to keep a generic PQ charge assignment, another option is to impose a slightly weaker, but equally efficient constraint, which is that the PQ symmetry is axial ``on average''\footnote{In the language of section \ref{section:GaugedAbelianMatching}, one does not need to demand that the PQ anomalous contribution of the GCS vanishes for each integrated massive fermion, but only that it does at the level of a subset or all of the heavy fermions.},
\be
X_{L_1} +X_{L_2} + X_{E_1}+ X_{E_2} = 0 \, , \qquad  X_{L_1} +X_{L_2} + X_{N_1}+ X_{N_2} = 0 \, .
\ee
This yields in particular
\be
\text{Axial PQ : } X_{L_1}+X_{L_2}=\frac{X_1-X_2}{2} \ ,
\ee
thanks to which the photon-$Z$ terms in \eqref{PQanomaliesAppendix} and \eqref{axionPhotonZcouplings} match. Finally, to understand the $Z\tilde Z$ coupling, let us first write down the kinetic terms in terms of photons and $Z$s:
\be
\bead
&e\overline{N_i}\gamma^\mu(-1)^{i+1}\left[\left(Y+\frac{1}{2}\right)A_\mu+\left(-t_W\left(Y+\frac{1}{4}+\frac{1}{4}\gamma_5\right)+t_W^{-1}\frac{1-\gamma_5}{4}\right)Z_\mu \right]N_i\\
&+e\overline{E_i}\gamma^\mu(-1)^{i+1}\left[\left(Y-\frac{1}{2}\right)A_\mu-\left(t_W\left(Y-\frac{1}{4}-\frac{1}{4}\gamma_5\right)+t_W^{-1}\frac{1-\gamma_5}{4}\right)Z_\mu \right]E_i\\
\eead
\ee
The photon coupling is of course vector-like. We know from \eqref{prescriptionFormula} or \eqref{prescriptionAppendix} that a systematic way to get the right $Z\tilde Z$ coupling from the anomaly is to choose the PQ symmetry such that $3\alpha_\text{PQ}\alpha_Z+\beta_\text{PQ}\beta_Z=0$ for each integrated field. That means demanding
\be
\bead
6X_{L_i}\left(-t_W\left(Y+\frac{1}{4}\right)+\frac{t_W^{-1}}{4}\right)+(-1)^{i}X_{i+1\text{ mod } 2}\left(t_W\left(3Y+1\right)-\frac{t_W^{-1}}{2}\right)=0\\
6X_{L_i}\left(-t_W\left(-Y+\frac{1}{4}\right)+\frac{t_W^{-1}}{4}\right)+(-1)^{i+1}X_{i}\left(t_W\left(-3Y+1\right)-\frac{t_W^{-1}}{2}\right)=0 \ .
\eead
\label{conditionsZZ}
\ee
Those conditions are again too restrictive, namely they impose conditions on $X_{1,2}$, but choosing $\sum(3\alpha_\text{PQ}\alpha_Z+\beta_\text{PQ}\beta_Z)=0$ for a whole ($N_i,E_i$) pair is allowed for generic Higgs charges, and sufficient. The condition to enforce is the sum of the two contributions in \eqref{conditionsZZ},
\be
3X_{L_i}(t_W^{-1}-t_W)+(-1)^iX_{i+1\text{ mod } 2}\left(t_W\left(3Y+1\right)-\frac{t_W^{-1}}{2}\right)+(-1)^{i+1}X_i\left(t_W\left(-3Y+1\right)-\frac{t_W^{-1}}{2}\right)=0 \ ,
\ee
hence
\be
X_{L_1}+X_{L_2}=\frac{t_W^{-1}-2t_W}{3(t_W^{-1}-t_W)}(X_1-X_2)
\ee
With this choice, the $aZ\tilde Z$ in \eqref{PQanomaliesAppendix} and \eqref{axionPhotonZcouplings} match.

The couplings to the charged bosons $W^{\pm}$ can also be understood along those lines. Comparing \eqref{PQanomaliesAppendix} and \eqref{axionPhotonZcouplings}, we see that we would like that
\be
X_{L_1}+X_{L_2}=\frac{X_1-X_2}{3}
\label{constraintWW}
\ee
for them to match. This relation is again achieved when we enforce that $3\alpha_\text{PQ}\alpha_W+\beta_\text{PQ}\beta_W=0$, where $\alpha_W=\beta_W$ since the coupling to $W$s is purely left-handed. We cannot enforce it at the level of each fermion without constraining $X_{1,2}$, as we are now quite used to, but we can impose a similar constraint on a full doublet ($N_i,E_i$). It means that $4X_{L_i}+X_{N_i}+X_{E_i}=0$. With \eqref{PQrelations}, we see that this gives \eqref{constraintWW}.

As a final remark, notice that we did not need to worry about the PQ variation of other hypothetical axion terms, such as the (pure gauge) ones that feature the longitudinal component of the $Z$ boson $a_Z$, of the form $a_Z F\tilde F$, etc. This is due to the fact that we chose the PQ symmetry so that $a_Z$ is PQ neutral, as can be seen from \eqref{eq:DFSZHiggsX}.



\bibliographystyle{apsrev4-1_title}
\bibliography{biblio.bib}


\end{document}